\documentclass[article, shortnames]{jss}

\usepackage{orcidlink,thumbpdf,lmodern}
\usepackage{amsmath, amsfonts, amssymb}
\usepackage{algorithm, algpseudocode}     
\usepackage{float}

\usepackage{framed}

\author{Juliette Ortholand\\Sorbonne Université,\\Amsterdam UMC
       \And Sofia Kaisaridi\\Sorbonne Université
       \And Nicolas Gensollen\\Sorbonne Université,\\Univ. Bordeaux
       \And Etienne Maheux\\Sorbonne Université 
       \AND Caglayan Tuna\\Sorbonne Université,\\Université Paris Cité
       \And Raphael Couronne\\Sorbonne Université 
       \And Arnaud Valladier\\Sorbonne Université 
       \AND Pierre-Emmanuel Poulet\\Sorbonne Université 
       \And Nemo Fournier\\Sorbonne Université 
       \And Léa Aguilhon\\Sorbonne Université 
       \AND Maylis Tran\\Sorbonne Université  
       \And Gabrielle Casimiro\\Sorbonne Université 
       \And Jean-Vincent Martini\\ Université Paris-Saclay
       \AND Sebastian Mendez\\Sorbonne Université,\\Université Paris Cité
       \And Igor Koval\\Sorbonne Université
       \And Stanley Durrleman\\Sorbonne Université,\\Qairnel SAS
       \AND Sophie Tezenas Du Montcel\\Sorbonne Université 
}
\Plainauthor{Juliette Ortholand, Sofia Kaisaridi, Nicolas Gensollen, Etienne Maheux, Caglayan Tuna, Raphael Couronne, Arnaud Valladier, Pierre-Emmanuel Poulet, Nemo Fournier, Léa Aguilhon, Maylis Tran, Gabrielle Casimiro, Sebastian Mendez, Igor Koval, Stanley Durrleman, Sophie Tezenas Du Montcel}

\title{leaspy: LEArning Spatiotemporal Patterns in PYthon}
\Plaintitle{leaspy: LEArning Spatiotemporal Patterns in PYthon}
\Shorttitle{leaspy: LEArning Spatiotemporal Patterns in PYthon}
\Abstract{
  Longitudinal data are fundamental across scientific disciplines for modeling how complex systems evolve over time. A core challenge in these settings is handling temporal misalignment: different subjects undergo a similar underlying process but at varying speeds and starting times. This difficulty is further compounded when tracking multivariate dynamics, where features interact dynamically rather than following simple, independent pathways. To address these challenges, we present \pkg{leaspy} (LEArning Spatiotemporal patterns in PYthon), an open-source \proglang{Python} library. Built on a mixed effects model, \pkg{leaspy} enables the estimation of population-level trajectories while accounting for subject-specific variability. The library supports multivariate formulation across diverse data types, including continuous, time-to-event (joint), and mixture models-and has been successfully applied to characterize disease heterogeneity, and generate individual predictions We demonstrate its practical utility through an application in neurodegenerative disease progression. Developed following modern software engineering practices, including systematic testing and continuous integration, \pkg{leaspy} facilitates the integration of new models and provides a robust user-friendly library for longitudinal progression modeling.
}

\Plainkeywords{multivariate trajectories, longitudinal, mixed-effects, Python}

\Address{
  Juliette Ortholand\\
  ARAMIS, Institut du Cerveau-Paris Brain Institute\\
  Sorbonne Université, CNRS, Inria, Inserm,\\ 
  AP-HP, Groupe Hospitalier Sorbonne Université,\\
  Paris, France \\
  \emph{and}\\
  KIK, Amsterdam UMC, \\
  Amsterdam, Netherlands\\
  E-mail: \email{j.m.ortholand@amsterdamumc.nl}\\

  Sofia Kaisaridi, Sophie Tezenas du Montcel\\
  ARAMIS, Institut du Cerveau-Paris Brain Institute\\
  Sorbonne Université, CNRS, Inria, Inserm,\\ 
  AP-HP, Groupe Hospitalier Sorbonne Université,\\
  Paris, France \\
  E-mail: \email{sofia.kaisaridi@icm-institute.org}, \email{sophie.tezenas@aphp.fr}
}

\begin{document}



\section{Introduction}

Longitudinal and repeated-measurement data are fundamental across diverse scientific disciplines for modeling the trajectories of complex systems as they evolve over time. 
A core statistical challenge in analyzing these evolutionary pathways is handling inherent unaligned observations: different subjects often undergo a similar underlying process, but at varying speeds and distinct starting times.
A common statistical framework for analyzing such data is the class of generalized linear mixed-effects models (GLMMs), which enables the estimation of population-level trajectories while accounting for subject-specific variability through random effects ~(\cite{rizopoulos_r_2016, proust-lima_estimation_2017, rondeau_frailtypack_2012, seabold2010statsmodels}). 
These models have become standard tools in practice and are implemented in several statistical libraries, including \pkg{statsmodels}~(\cite{seabold2010statsmodels}) and \pkg{GLMMadaptive}~(\cite{glmm}).
However, complex systems often exhibit intricate temporal dynamics that cannot be adequately captured by linear or generalized linear frameworks. 
To capture these complex behaviors, researchers frequently turn to the broader class of nonlinear mixed-effects models (NLMEMs), implemented in powerful specialized frameworks such as \pkg{saemix}~(\cite{comets_parameter_2017}).

A key assumption of most standard GLMM and NLMEM approaches is that observations can be aligned relative to a known reference time, such as an onset, a critical event, or study baseline~(\cite{schiratti_bayesian_2017}). 
This temporal misalignment problem arises across diverse domains, such as economics (e.g., market life cycles) and engineering (e.g., mechanical degradation), and has been most prominently studied within the framework of biomedical disease progression modeling. 
In these contexts, a universal reference point is frequently unobserved or poorly defined. Individuals may enter a study at different latent stages of an underlying process, making direct temporal alignment challenging and potentially biasing trajectory estimates. To address this limitation, a class of progression models employing data-driven time reparametrization has emerged that seeks to infer latent temporal structures directly from the data~(\cite{young_data-driven_2024}). 
Early approaches relied on ordinary differential equation (ODE) models to describe feature dynamics~(\cite{samtani_disease_2013, villemagne_amyloid_2013, jack_brain_2013, oxtoby_learning_2014}).
Although having some promising results, they lack a clear separation of individual and population variability.
Nonparametric methods based on Gaussian processes have also been introduced (\pkg{GP\_progression\_model}), providing flexible estimates of population-level trajectories~(\cite{abi_nader_monotonic_2020}). 
Other methods infer progression from cross-sectional data, such as the \pkg{SuStaIn} framework~(\cite{aksman_pysustain_2021}) which models progression as a sequence of discrete events and can identify distinct subtypes. 
To our knowledge, it has not been extended to a longitudinal framework to this day.
To combine latent temporal alignment with mixed-effects modeling, several approaches introduce subject-specific latent variables, such as a progression score or a process age~(\cite{jedynak_computational_2012, bilgel_multivariate_2016, bilgel_predicting_2019,li_bayesian_2017, iddi_estimating_2018}). 
For instance, the semi-parametric framework implemented in the \pkg{grace} package~(\cite{donohue_estimating_2014}) offers high flexibility by simultaneously estimating non-parametric curves alongside subject-specific temporal ordering, but relies on the abundance of available data to ensure convergence. 
Alternative non-linear mixed-effects models (NLMEMs), such as \pkg{progmod}~(\cite{raket_statistical_2020}), map calendar time to an intrinsic timeline modeling individual trajectory variations through localized stochastic components. 
While these implementations are methodologically powerful, they often rely on iterative optimization shortcuts or traditional analytical approximations that can face structural convergence obstacles when dealing with sparse, highly non-linear, or high-dimensional multivariate tracking data.
More recently, some NLMEM frameworks have integrated observable clinical milestones into their formulation, anchoring individual latent time-shifts around specific diagnostic events~(\cite{lespinasse_2023}). 
While this enhances interpretability, requiring an observable reference point is a distinct limitation since such milestones are often unavailable in practice.

Capturing multivariate dynamics is also critical, as the progression of complex systems typically involves multiple interacting outcomes. 
Classical multivariate models often rely on Euclidean assumptions that may fail to capture intricate interactions between these variables. 
In contrast, geometric approaches provide a principled framework for modeling trajectories in nonlinear spaces while preserving the intrinsic structure of the data, offering a more flexible and accurate representation of the evolution dynamics of different feature domains.
This perspective builds on statistical shape analysis and computational anatomy~(\cite{Grenander1998, Miller2006, Allassonniere2007, Allassonniere2010})
and has proven highly effective for modeling the trajectories of complex, high-dimensional processes that exhibit heterogeneous and multidimensional progression.
While particularly relevant in neurodegenerative and chronic diseases, where tracking diverse, asynchronous biomarkers is critical, this geometric paradigm is equally applicable to non-medical domains. 
For instance, similar challenges arise when tracking multi-sensor industrial degradation patterns, environmental climate trajectories, or evolving economic indicator pathways, where multiple interacting features must be aligned across entities moving at different speeds.

A progression model that integrates a time reparametrization scheme and a meaningful multivariate representation has been introduced as the Disease Course Mapping (DCM) model~(\cite{Schiratti2015, Schiratti2017}).
It is formulated within a Riemannian geometry framework, offering a principled way to capture feature-specific differences for each individual. 
Individual trajectories are aligned along a common, latent timeline through subject-specific random effects capturing variability in both the process onset and its progression rate.
Originally developed for continuous trajectories, the framework has been extended to incorporate joint modeling with time-to-event outcomes ~(\cite{ortholand_joint_2024, ortholand_joint_2025}), mixture models ~(\cite{poulet_mixture_2021, Kaisaridi2026}), and ordinal outcomes ~(\cite{poulet_multivariate_2023}).
It has already been proven useful in various settings such as the modeling of imaging data~(\cite{lorenzi_probabilistic_2019, wang_progression_2022, abi_nader_monotonic_2020}), discrete responses~(\cite{poulet_multivariate_2023, Moulaire2023}), and multivariate outcomes~(\cite{kaisaridi2025determining}). 
It has demonstrated strong predictive performance in neurodegenerative diseases, notably outperforming 56 competing methods in predicting cognitive decline in Alzheimer's disease during the TADPOLE challenge ~(\cite{marinescu_tadpole_2019, koval_ad_2021}).

The open-source \pkg{leaspy} library provides an accessible implementation of this progression modeling framework in \proglang{Python}. 
By bridging advanced statistical modeling with a practical and user-friendly interface, it enables applied researchers across various fields to leverage data-driven progression models for complex longitudinal data.
Ultimately, the package serves as a versatile tool for realigning, clustering, and predicting asynchronous trajectories, making sophisticated progression analysis highly accessible regardless of the specific application domain.
The remainder of the paper is structured as follows. 
Section \ref{sec:models} introduces the theoretical background and mathematical formulation of the model. 
Section \ref{sect:implementation} describes key implementation choices and Section \ref{sect:main_functions} presents the main API functions. 
Section \ref{sec:illustrations} demonstrates the use of the package through a practical example in neurodegenerative disease progression modeling.
Section \ref{sec:discussion} discusses the contributions presented and provides general conclusions.



\section{Theoretical background} \label{sec:models}

\subsection{Modeling}

We consider $N$ subjects undergoing $n_i$ visits, with $t_{ij}$ denoting the age of subject $i$ $(i = 1, \dots, N)$ at visit $j$ $(j = 1, \dots, n_i)$.  
At each visit, up to $d$ features are assessed, with $y_{ijk}$ representing the value of feature $k$ $(k = 1, \dots, d)$ for subject $i$ at visit $j$. 

\subsubsection{Riemannian framework}

The proposed progression model was originally formulated within a Riemannian geometry framework.
A full understanding of the mathematical background is not required to follow the model and its use in practice; interested readers can find comprehensive derivations in previous work~(\cite{Koval_thesis, Schiratti2015, Schiratti2017}). 
We nevertheless provide an intuitive overview of the model and the rationale behind its construction.

To give geometrical intuition, we first detail a univariate case where observations denoted by $y_{ijk}$, are assumed to lie on an one-dimensional Riemannian manifold $\mathcal{M}=(0,1)$, an open interval of the real line, that locally resembles Euclidean space.
It is equipped with a Riemannian metric $g$, a function that defines inner products on tangent spaces at each point. 
Conceptually, these tangent spaces represent all possible directions of movement at a given point, and the inner product is a function that measures the length and angle of these directions.
Since our manifold is 1-dimensional, meaning we can move only in one direction, the tangent space is also 1-dimensional.

\begin{figure}[h]
    \centering
    \includegraphics[width=0.5\linewidth]{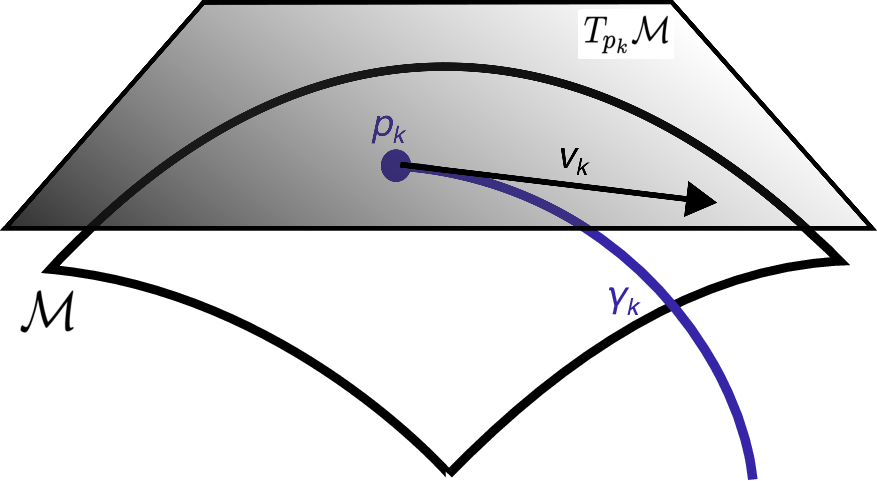}
    \caption{A curved surface represented as a manifold $\mathcal{M}$ and the geodesic $\gamma_k$ for the $k$-th score that passes from a point $p_k$ with velocity $v_k$. $T_{p_k}\mathcal{M}$ is the associated tangent space of the manifold $\mathcal{M}$ on the point $p_k$.}
    \label{fig: manifold_tangent}
\end{figure}

We consider a point $p_k$ in the manifold $\mathcal{M}$, a velocity $v_k$ in the tangent space $T_{p_k}\mathcal{M}$, and a geodesic $\gamma_k$, which is a smooth curve that locally minimizes distance according to the metric.
This curve corresponds to the latent trajectory, which at a given time point $t_0$ passes through the point $p_k$, $\gamma_k(t_0) = p_k$, with velocity of $v_k$, and can be interpreted as the progression dynamics over time $t$.
The shape of this trajectory is determined by the choice of the Riemannian metric applied to the manifold. 
For example, the Riemannian metric $g_p(u,v)=\frac{uv}{p^2(1-p)^2}$ corresponds to a logistic progression.
In a multivariate context, each score is associated with a 1D manifold, which is then put together as a product of 1D manifolds. 
The metric built over this space is defined as the product of the metrics of each 1D manifold.

When modeling clinical data, such as those analyzed in our illustration, floor and ceiling effects are common considerations, making the logistic curve a natural choice.
Its characteristic s-shaped pattern reflects the typical evolutionary dynamics of many progressive processes, with the inflection point capturing the critical midpoint of maximum acceleration before the system ultimately approaches its natural limits.
When modeling biomarkers or other unconstrained data modalities lacking distinct floor and ceiling effects, the logistic function remains versatile, adjusting to capture near-linear trajectories. 
For the rest of this paper, we will use the multivariate logistic model as a working example to illustrate the relevant equations. 
A linear version is also available in \pkg{leaspy}, and the framework is designed to allow developers to define new shapes as needed.

\subsubsection{Mixed-effects model: Latent common timeline and dimension reduction} \label{sect:method_random_effect}

Building on a mixed-effects formulation, the Disease Course Mapping model employs a nonlinear structure with time reparameterization to disentangle inter- and intra-patient variability in disease progression.
The normalized observations $y_{ijk}$ are then modeled as:

\begin{eqnarray}
y_{ijk} = \left(1 + (\frac{1}{p_k}-1)exp(-\frac{v_k(e^{\xi_i}(t_{ij}-\tau_i))+w_{ik}}{p_k(1-p_k)})\right)^{-1} +\epsilon_{ijk} \label{eq: DCM_statmodel}
\end{eqnarray}

Following the hierarchical structure commonly used in mixed-effects models, two types of parameters are introduced to capture both population- and individual-level effects.
The population-level parameters $p_k$ and $v_k$ represent, respectively, the coordinate of the position and velocity (i.e. the derivative of the curve) for the $k$-th feature at the inflection point, which in the context of disease progression modeling represents the average disease onset time $\overline\tau$.
The individual-level parameters $\xi_i$ and $\tau_i$ capture inter-individual temporal variability, while $w_{ik}$ accounts for spatial variability across subjects.
Finally, a Gaussian noise term is added to account for measurement variability $\epsilon_{ijk} \sim \mathcal{N}(\mathbf{O}_d, \sigma_k^2\mathbf{I}_d)$.

\paragraph{Latent common timeline:}

The time reparametrization function $e^{\xi_i}(t_{ij}-\tau_i)$ maps an individual’s time of measurement to a latent, intrinsic timeline, with $\xi_i$ and $\tau_i$ denoting the individual-specific progression rate and onset time.
While this latent timeline is often interpreted as a "disease age" within biomedical applications, it serves as a general-purpose temporal realignment applicable to any setting where subjects transition through a common process, each exhibiting a unique evolutionary profile.
This formulation enables comparisons at equivalent positions along the progression trajectory rather than at similar measurement times.
For instance, two patients with different chronological ages, such as a 40 year old with early onset and slow progression, and a 55 year old with later onset but faster progression, may pass by the same stage of the disease at a specific time $t'$.
The model therefore maps them at time $t'$ to an identical latent age, allowing their score trajectories to be compared on a shared latent disease timeline rather than on calendar time.

\paragraph{Spatial random effects:}

The variables $w_{ik}$ capture spatial variability, in a topological rather than a geographic sense, and can be interpreted as inter-score spacing parameters.
To reduce computational complexity and ensure identifiability, rather than being estimated directly, they are expressed through an ICA-like decomposition $\mathbf{w_i} = \mathbf{A}\cdot \mathbf{s_i}$, where $\mathbf{A}$ denotes a mixing matrix and $\mathbf{s_i}$ is a vector of latent parameters, referred to as sources.
The number of sources $N_s$ is chosen to be smaller than the total number of features $d$ and it is generally recommended to select $N_s \approx \sqrt{d}$. 
The mixing matrix  $\mathbf{A} \in \mathbb{R}^{d\times N_s}$ is an intermediate (linked) variable composed of $N_s$ columns that are orthogonal to the population velocity vector $\mathbf{v}$ to ensure identifiability.
It is obtained by constructing an orthogonal basis $(\mathcal{B}_m)_{1 \leq m \leq d-1}$ of $Span(\mathbf{v})$ using the Householder method~(\cite{Householder1958}). 
Each column of $\mathbf{A}$ is then expressed as a linear combination of the basis vectors and the population parameters $\{\beta_{ml}\}_{1 \le m \le d-1,\; 1 \le l \le N_s}$ so that: $\mathbf{A_{:,l}} = \sum_{m=1}^{d-1} \beta_{ml}\mathcal{B}_m$.

\subsubsection{Adaptation to various data types}

To accommodate various types of observational data corresponding to different situations, \pkg{leaspy} enables the implementation of new models.
In its current version, two additional model classes are provided: the \code{joint} model~(\cite{ortholand_joint_2024}) and the \code{mixture} model~(\cite{Kaisaridi2026}).

\paragraph{Joint model:}
Joint models are a class of statistical models that simultaneously analyze longitudinal and censored data~(\cite{Alsefri2020, Ibrahim2010}).
Unlike traditional approaches that treat these processes separately, joint models integrate them into a unified framework, recognizing that they often share underlying mechanisms—for example, a slowly progressing biomarker may signal an increased risk of a clinical event. 
By linking the two submodels—typically, here through shared random effects~(\cite{Rizopoulos2011}), the models account for their interdependence, reducing biases from informative dropout or measurement error.

\paragraph{Mixture model:}
Mixture models are a class of statistical models that represent a population as a combination of underlying subpopulations, each described by its own probability distribution~(\cite{McLachlan2000}). 
Standard mixed-effects models assume a certain homogeneity in the sense that inter-patient variability is considered random variation around a fixed reference. 
Mixture models explicitly recognize that observed data may arise from distinct latent groups—for example, subjects may cluster into distinct subtypes based on shared characteristics, such as similar onset timings, rates of progression, or overall trajectory profiles. 
This formulation captures heterogeneity in the data, allowing for flexible modeling of complex, multimodal patterns. 

\subsection{Estimation and prediction} \label{subsec: est_pred}

In nonlinear mixed-effects models, the observed likelihood does not admit a closed-form expression. 
Consequently, an analytical maximization of the log-likelihood is not feasible. 
To estimate the model parameters, we employ a stochastic version of the Expectation–Maximization (EM) algorithm, namely the MCMC-SAEM algorithm (Markov Chain Monte Carlo – Stochastic Approximation Expectation Maximization). For models belonging to the family of curved exponential functions ~(\cite{Kuhn2004})—which includes all the models studied in this work—the stochastic approximation mechanism guarantees convergence.

\subsubsection{Characterization of parameters}
The model involves three categories of parameters with distinct roles in the estimation process: latent parameters $z$, model parameters $\theta$, and hyperparameters $\Pi$.
To remain consistent with the internal organization of \pkg{leaspy}, we describe the categories as implemented in the codebase defined by the role during the fitting procedure. 
Drawing a parallel to the standard mixed-effects models formulation, the latent individual parameters represent the random effects of the model, while the model parameters represent the fixed effects.

\begin{itemize}
    \item[$\bullet$] latent parameters: $\mathbf{z} =\{\mathbf{z}_{pop}, (\mathbf{z}_{re,i})_{1 \leq i \leq N}\}$ 
    
    Latent parameters are sampled during fitting according to probability distributions that are defined directly or after transformations ($g_k = \frac{1 - p_k}{p_k}, \quad \tilde{g}_k = \log (g_k), \quad \tilde{v}_k = \log (v_k)$). We distinguish between population parameters $\mathbf{z}_{pop} =(\tilde{g_k}, \tilde{v_k}, \beta_{ml})$ which involve the whole sample to be updated during estimation, and individual parameters $\mathbf{z}_{re,i} =(\tau_i, \xi_i, s_{il})$ which only require data points of individual $i$ to be updated.
    They are assumed to follow Gaussian distributions defined by the model parameters, as described below. The assumption of priors on the population parameters serves a regularizing purpose rather than dictating bayesian inference. This approach penalizes overparameterization, thereby protecting the stability of the global population trajectory from being distorted by individual-specific noise.

\begin{equation}
\begin{minipage}[t]{0.48\linewidth}
\centering
\textbf{Population-level parameters:}
\[
\begin{cases}
    \tilde{g}_k \sim \mathcal{N}(\overline{\tilde{g}_k}, \sigma^2_{\tilde{g}}),\\[4pt]
    \tilde{v}_k \sim \mathcal{N}(\overline{\tilde{v}_k}, \sigma^2_{\tilde{v}}),\\[4pt]
    \beta_{ml} \sim \mathcal{N}(\overline{\beta_{ml}}, \sigma^2_{\beta})
\end{cases}
\]
\end{minipage}
\hfill
\begin{minipage}[t]{0.48\linewidth}
\centering
\textbf{Individual-level parameters:}
\[
\begin{cases}
    \xi_i \sim \mathcal{N}(\overline{\xi}, \sigma^2_{\xi})\\
    \tau_i \sim \mathcal{N}(\overline{\tau}, \sigma^2_{\tau})\\
    s_{il} \sim \mathcal{N}(\overline{s},\sigma^2_{s})
\end{cases}
\]
\end{minipage}
\label{eq : latent_params}
\end{equation}

    It is worth noting that directly estimating the population parameters can lead to identifiability issues due to the strong mathematical coupling between the global parameters (defining the average trajectory) and the individual random effects (defining individual deformations).
    We also risk overfitting the noise, leading to a degenerate solution where the population trajectory loses its structural meaning. 
    To address this employ an hierarchical architecture, enforcing priors with small variances for the population parameters.
    This acts as a regularization mechanism, penalizing overparameterization and successfully decoupling the population-level trends from individual-specific variability.
    
\end{itemize}

\begin{itemize}
    \item[$\bullet$] model parameters: $\theta =\{\overline{\tilde{g}}_k,\overline{\tilde{v}}_k, \overline{\beta}_{ml}, \overline{\tau}, \sigma_{\tau}, \sigma_{\xi}, \sigma_{k}\}$
    
    These constitute the set of parameters to be estimated. For the population-level parameters, the relevant estimates correspond to the mean values of the distributions of $(\tilde{g_k}, \tilde{v_k}, \beta_{ml})$. These values are used as the optimal estimates within the model. For the temporal individual parameters $(\tau_i, \xi_i)$ they correspond to the mean and/or the standard deviation of the Gaussian distribution associated with these random effects. Finally the term $\sigma_{k}$ characterizes the standard deviation of the Gaussian noise added to the observations.
    
\end{itemize}

\begin{itemize}
    \item[$\bullet$] hyperparameters: $\Pi = \{\sigma_{\tilde{g}}, \sigma_{\tilde{v}}, \sigma_{\beta}, \overline{\xi}, \overline{s}, \sigma_s\}$
    
    Their values are chosen in advance. These include $\sigma_{\tilde{g}}$, $\sigma_{\tilde{v}}$, $\sigma_{\beta}$ which are all fixed to 0.01 by default and appear in priors involving the same scheme: $x \sim \mathcal{N}(\overline{x}, \sigma^2_x)$. However, these priors do not influence the posterior distribution, as they initialize the stochastic exploration of the parameter space during the simulation step. During the algorithm, these standard deviations are updated to maintain an optimal acceptance rate. The set of hyperparameters also includes $\overline{\xi}$ set to 0 corresponding to the mean of the random effects $\xi_i$. We also fix the parameters describing the Gaussian distribution of the sources. For identifiability reasons we set $\overline{\xi}=0, \overline{s}=0, \sigma_s=1$.
\end{itemize}

\subsubsection{Fit algorithm}

The MCMC-SAEM algorithm allows for maximum likelihood estimation in the presence of latent variables when the observed likelihood is not available in closed form.
In practice, the algorithm alternates between three steps:

\begin{enumerate}
    \item a simulation step, where the latent variables $\mathbf{z}$ are sampled according to their distributions, given the current estimates of the model parameters using a Metropolis-Hastings algorithm with a Gibbs sampler;
    \item a stochastic approximation step, where instead of computing the expected complete-data log-likelihood explicitly, its sufficient statistics are updated recursively using a stochastic averaging scheme~(\cite{Robbins1951});
    \item a maximization step, where the model parameters $\boldsymbol{\theta}$ are updated by maximizing the approximate expected complete-data log-likelihood.
\end{enumerate}

\subsubsection{Personalize algorithm}

Once the parameters of the model are estimated, a personalization algorithm can be employed to obtain reliable estimates or predictions of the random effects, conditional on the previously estimated fixed effects. 
By estimating these individual parameters, \pkg{leaspy} characterizes each subject’s trajectory and enables predictions of their future progression.
Two main approaches are available to perform this step.

From a frequentist perspective, the random effects are estimated conditionally on the fixed effects by maximizing the individual likelihood using numerical optimization. 
This optimization is performed with the minimize procedure from the \pkg{SciPy} library, which provides a unified interface to several gradient-based and derivative-free algorithms. 
In our implementation, the objective function corresponds to the negative log-likelihood (NLL) of the individual parameters given the current estimates of the fixed effects. 
This approach yields maximum likelihood estimates of the random effects, conditional on the fixed effects, for each individual.

From a Bayesian perspective, the Gibbs sampler used for the MCMC-SAEM can also be used to generate draws from the conditional posterior distribution of the individual parameters, given the current values of the model parameters and the observed data. 
The package currently provides two summary estimates of these posterior distributions: the posterior means and the posterior modes.

Because longitudinal mixed-effects models require a strict hierarchical data structure composed of subjects with multiple repeated measures, the framework is not compatible with standard \pkg{scikit-learn} cross-validation methods, which assume independent and tabular observations. While a subject-level K-fold cross-validation of the population-fitting step could in principle be implemented manually, it was not a primary design goal of the package and incurs a substantial computational cost, as the MCMC-SAEM fitting procedure would need to be repeated K times. Cross-validation is more naturally suited to the personalization step: individual parameters are estimated from a subset of a subject's visits, and the held-out visit is then predicted and compared to the ground truth, providing an unbiased estimate of the model's predictive performance. This leave-one-visit-out evaluation, illustrated in the prediction section of the illustration, can be repeated across subjects to aggregate prediction metrics such as MAE or R2.

\subsubsection{Simulate algorithm}

The simulation algorithm operates through the following steps:

\begin{enumerate}
    \item A random sampling of the individual parameters step, where the individual-level parameters (xi, tau, and source components) are drawn from a Gaussian distribution specified by the user-defined model parameters. Space-shift values are then derived from the source components and the mixing matrix.
    \item  A visit timepoints generation step, where the baseline ages and follow-up visit timepoints are generated given user-specified settings (sampling from a Gaussian distribution, pre-defined timepoints dataframe...)
    \item A score estimation step, where the estimate algorithm is applied to compute the score values for each individual based on their sampled parameters.
    \item A noise injection step, where a beta-distributed noise is added to the simulated score values to introduce realistic variability.
\end{enumerate}

\section{Implementation}\label{sect:implementation}

\pkg{leaspy} occupies a distinct position within the software ecosystem for Nonlinear Mixed-Effects Models (NLMEMs), offering a dedicated \proglang{Python}-based alternative to established statistical environments. 
Traditional frequentist frameworks, such as \pkg{nlme}~(\cite{nlme}) and \pkg{lme4}~(\cite{lme4}) in \proglang{R} or \pkg{statsmodels}~(\cite{seabold2010statsmodels}) in \proglang{Python}, heavily rely on analytical approximations, using Laplacian or first-order conditional estimation (FOCE) methods, which can struggle to converge when modeling highly non-linear, multi-dimensional individual trajectories. 
In contrast, advanced \proglang{R} implementations like \pkg{saemix} ~(\cite{comets_parameter_2017}) employ the SAEM algorithm, which leverages an MCMC-based simulation step to ensure robust global convergence without requiring analytical linearization. 
On the fully Bayesian end of the spectrum, packages like \pkg{brms}~(\cite{brms}) rely on computationally intensive Hamiltonian Monte Carlo (HMC) samplers via Stan. 
Our framework bridges a crucial gap by addressing both usability and architectural limitations found in existing software. 
While frameworks like \pkg{saemix} require researchers to manually hardcode the structural model, rendering them highly flexible but largely destined for expert statisticians, \pkg{leaspy} provides accessible functionalities tailored for broader applications.
Furthermore, it is built on a robust, class-based object-oriented architecture with modern continuous integration and continuous deployment (CI/CD) standards in mind.
This software design deliberately bridges the gap between methodological developers needing structural stability and clinical researchers requiring reliable, automated deployment in practical settings.

The \pkg{leaspy} library is developed with a strong focus on code quality, adhering to the \code{ruff} style guide and maintaining a minimal set of dependencies. 
To ensure robustness and long-term maintainability, the library integrates Continuous Integration (CI) pipelines on GitHub, which automatically execute a comprehensive suite of unit tests covering all core functionalities. 
The documentation is generated using \code{Sphinx}, and a dedicated website\footnote{\url{https://leaspy.readthedocs.io/en/stable/}} provides detailed guidance on installation, usage examples, and the underlying mathematical framework. 
Installation and dependency management are handled via \code{Poetry}, which simplifies environment setup and package distribution for both users and developers. 
This modular and transparent architecture ensures clarity, extensibility, and consistency, while its straightforward structure greatly simplifies the development and integration of new models or algorithm variants. 

\subsection{Architecture} 

\begin{figure}[H]
    \centering
    \includegraphics[width=0.9\linewidth]{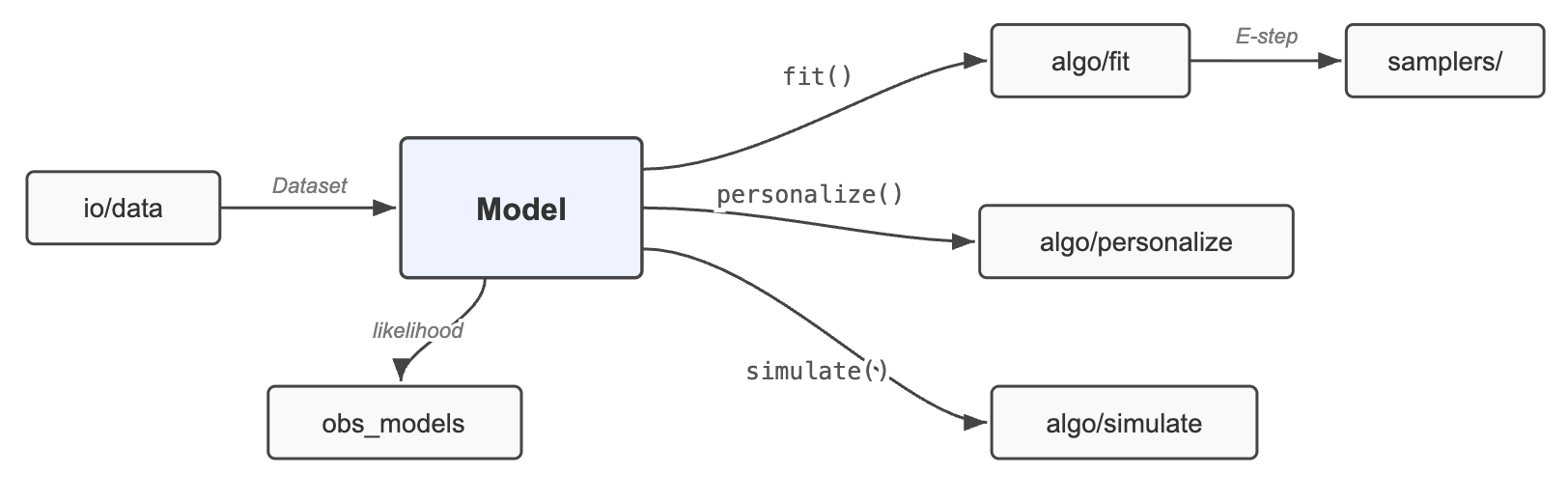}
    \caption{Overview of module interactions in the \pkg{leaspy} pipeline. The \code{Model} is the central entry point: \code{io/data} handles data loading, while \code{fit()}, \code{personalize()}, and \code{simulate()} dispatch to their respective algorithm modules. The sampler used during the E-step is configurable via \code{samplers/}, and \code{obs\_models} defines the computation of the NLL. Each module/model can be added or modified independently.}
    \label{fig: module}
\end{figure}

The library is built in a modular way: most methods are encapsulated in classes responsible for specific parts of the engine. 
Depending on the model and algorithm chosen, \pkg{leaspy} selects and instantiates the appropriate classes at runtime. 
The model interacts with other modules at each stage of the pipeline to realize the different tasks detailed in section \ref{subsec: est_pred}: the \code{io/data} module handles data loading and validation; calling \code{fit()} invokes classes in \code{algo/fit}; calling \code{personalize()} relies on \code{algo/personalize}; and even within specific algorithmic steps, such as the E-step of the MCMC-SAEM algorithm, the sampler itself is configurable and defined in the \code{samplers/} module (Figure~\ref{fig: module}). 
API functions are further detailed in section \ref{sect:main_functions}.

This architecture invites developers to integrate new models or methods by branching into the appropriate part of the hierarchy, reusing most of the existing code. 
A key design principle is the separation between model definition and inference algorithm: models define the DAG and the likelihood, while algorithms operate generically on any compatible model. 
The instantiation of components is handled through factory functions (e.g., \code{algorithm\_factory}, \code{sampler\_factory}), so new components can be registered without modifying existing code. 
For instance, a new model can inherit from the existing base classes and leverage the MCMC-SAEM algorithm as-is; a custom sampler can be introduced for the E-step without modifying the rest of the pipeline; and new parameters can be incorporated by extending the DAG. 
This level of customization extends further: developers can define a custom geometry for the model manifold or implement a different time reparametrization scheme.

\subsection{Models}

\subsubsection{DAG}

\begin{figure}[H]
    \centering
    \includegraphics[width=1\linewidth]{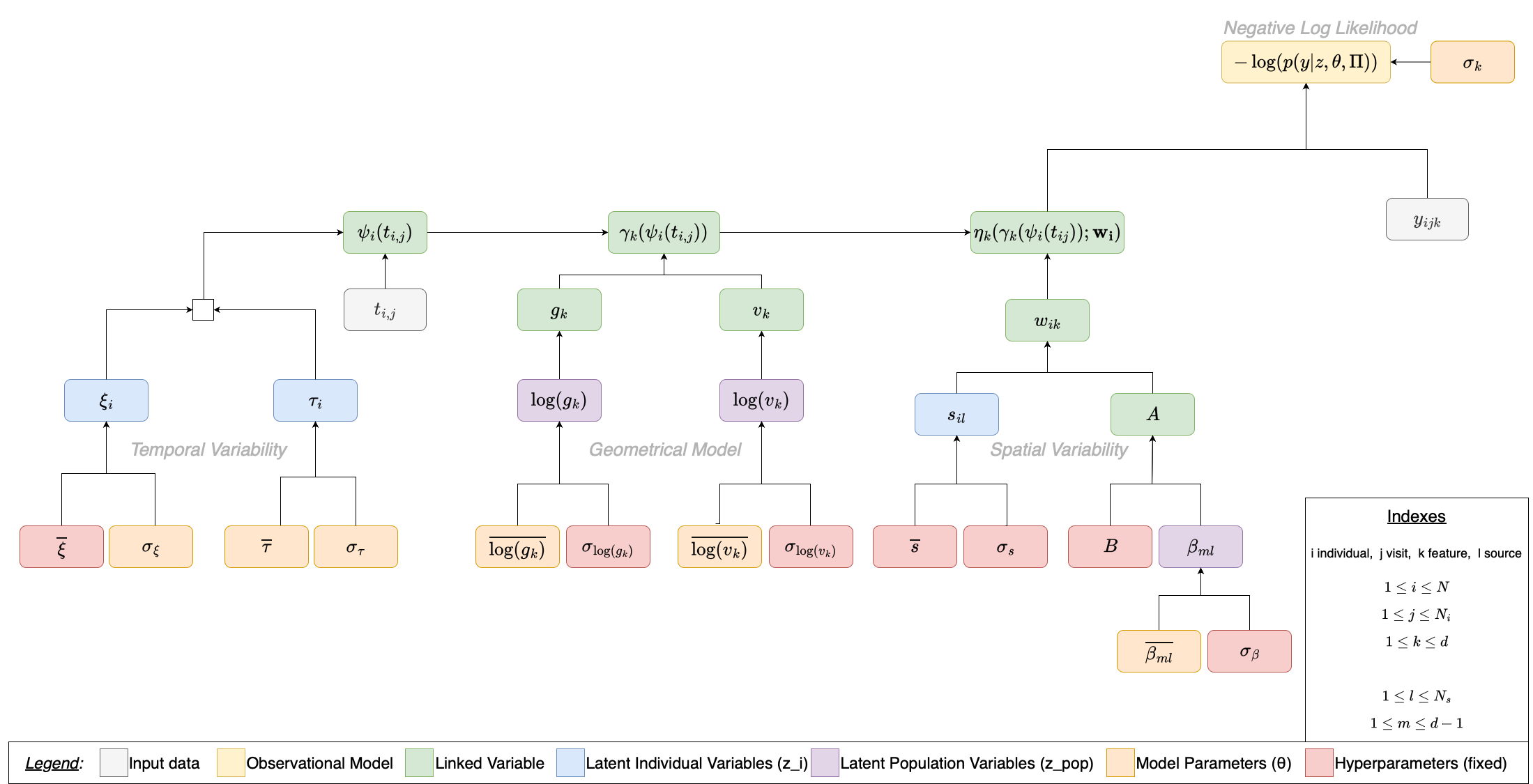}
    \caption{Directed acyclic graph (DAG) representing the multivariate logistic model. Colors indicate parameter categories, and arrows denote dependency relationships.}
    \label{fig: DAG}
\end{figure}

To optimize computations and keep the implementation close to the mathematical formulation, every model is defined with a Directed Acyclic Graph (DAG). 
The graph specifies the properties of each variable, the dependency structure, and the update rules. 
All the values of the variables associated with the DAG are stored in a \code{State} object. 
This structure was inspired by Bayesian software such as \pkg{PymC}~(\cite{pymc2023}). 
An example of a DAG representing the multivariate logistic model is presented in Figure~\ref{fig: DAG}. 
Different types of variables are stored in the DAG depending on the parameters they represent:

\begin{itemize}
    \item \code{DataVariable}: observed data (e.g.: $y_{ijk}$, $t_{ij}$),
    \item \code{Hyperparameter}: hyperparameters assigned fixed, predefined values (e.g.: $\overline{\xi}$),
    \item \code{ModelParameter}: parameters (fixed effects) that are maximized through the likelihood (e.g.: $\overline{\tau}$),
    \item \code{IndividualLatentVariable}: latent individual variables (random effects) that are sampled (e.g.: $\xi_i$, $\tau_i$),
    \item \code{PopulationLatentVariable}: latent population variables that are sampled (e.g.: $\log(g_k)$, $\log(v_k)$),
    \item \code{LinkedVariable}: variables obtained only from the computation of other variables (e.g.: $w_{ik}$), 
\end{itemize}

\subsubsection{Distribution Families} 

Following the ''family'' concept introduced in~\cite{rizopoulos2022package}, we define a class named \code{ObservationModel} that, given the model and observed data, specifies the linked variables associated with the observations and links them to the model through a well-defined distribution. 
Latent variables and observation models are based on probability distributions, implemented as specific classes that inherit from a common structure. 
The families currently implemented include the Gaussian distribution for longitudinal observations and the Weibull distribution with right censoring for event-time data.

\subsubsection{Implemented models} 

\begin{figure}[H]
    \centering
    \includegraphics[width=0.9\linewidth]{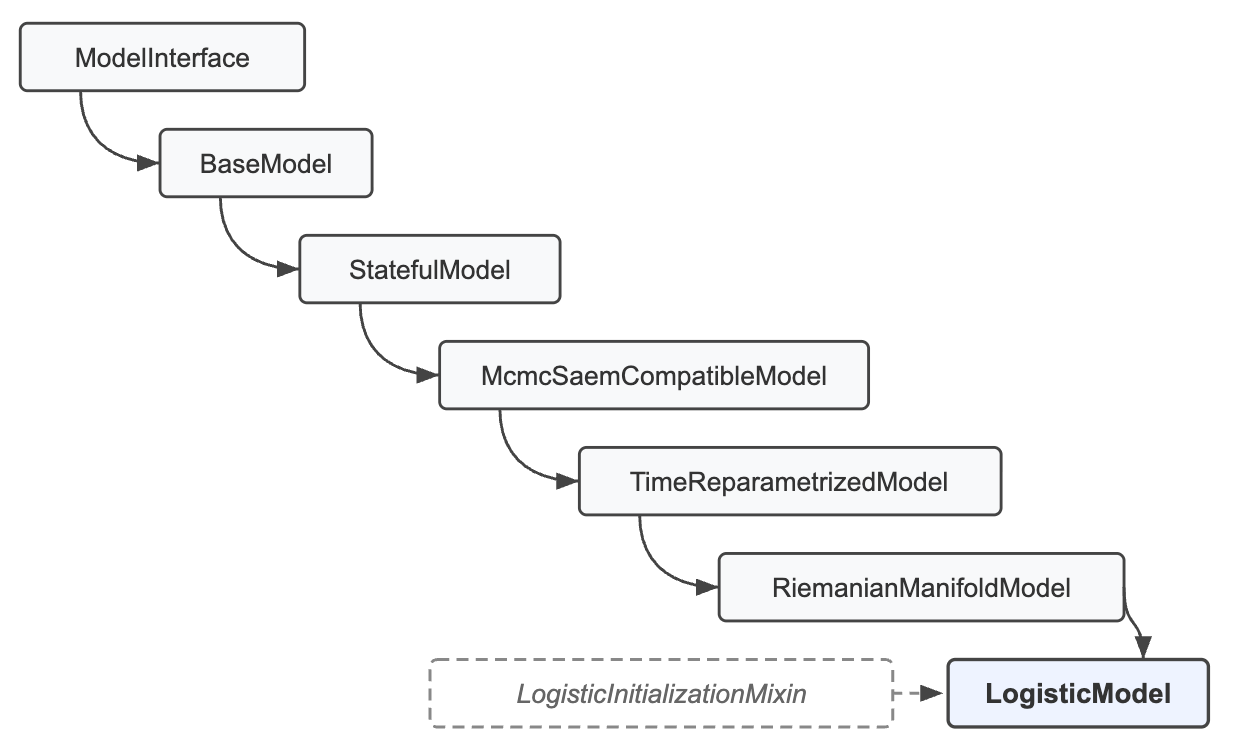}
    \caption{Inheritance chain each level of the hierarchy isolates a specific concern. \code{BaseModel} handles the data and runs the algorithm, \code{StatefulModel} adds the state tracking and handles the DAG, \code{McmcCompatibleModel} defines the MCMC-SAEM algorithm, \code{TimeReparametrizedModel} handles the time reparametrization, \code{RiemanianManifoldModel} implements the geometric framework. \code{LogisticinitializationMixin} initialize the parameters of \code{LogisticModel}, which contain the metric of the model.}
    \label{fig: inheritance}
\end{figure}

The three types of models implemented in \pkg{leaspy} are each defined by a dedicated class: the logistic Model (\code{LogisticModel}), the Joint Model (\code{JointModel}), and the mixture model (\code{LogisticMultivariateMixtureModel}). 
Taking \code{LogisticModel} as an example, it results from a chain of inheritance across several base classes (Figure~\ref{fig: inheritance}).

\section{API main functions} \label{sect:main_functions}

\begin{figure}[h]
    \centering
    \includegraphics[width=\columnwidth]{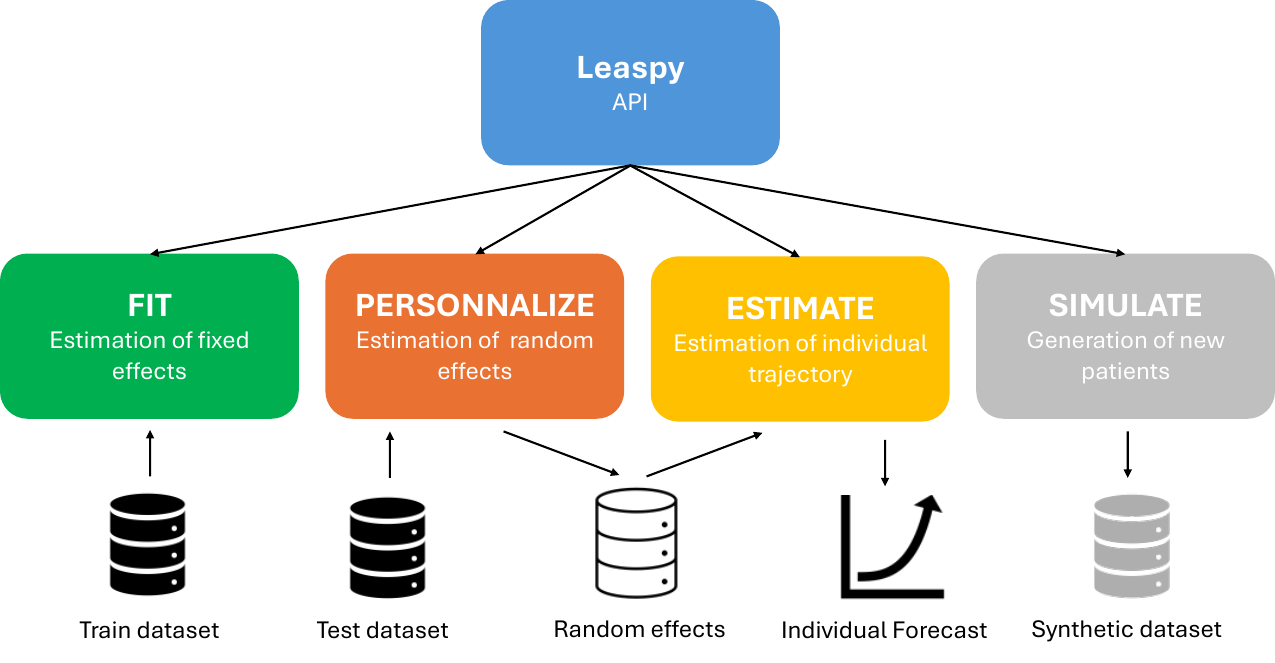}
    \caption{Overview of the main functionalities provided by the \pkg{leaspy} library.}
    \label{fig:leaspy}
\end{figure}

The longitudinal data, stored in a long format, can be processed using the main functions of the \pkg{leaspy} API to (i) estimate model parameters (\code{fit}), (ii) infer or predict individual progression (\code{personalize}, \code{estimate}), or (iii) simulate new datasets (\code{simulate}) from trained models, as illustrated in Figure~\ref{fig:leaspy}.

\subsection{Fit} 

\paragraph{What it does:}
The fit procedure estimates the model parameters (fixed effects) that characterize the population-level trends, by maximizing the total likelihood. 
This optimization is performed using the MCMC-SAEM algorithm (\cite{kuhn_coupling_2004}). 
Fitted \pkg{leaspy} models can be saved and reloaded using built-in functions, ensuring reproducibility and facilitating model sharing.

\paragraph{Arguments:} The main arguments of interest are described below.
\begin{itemize}
    \item \code{data}: The dataset to be used; it should follow \pkg{leaspy}'s formatting rules (futher specified in section \ref{sec:data_format}).
    \item \code{name}: The algorithm’s name. Currently, the only implemented algorithm is \code{'mcmc\_saem'}.
     \item \code{seed}: If provided, the seed initializes the random number generator so it can produce the same sequence of numbers, ensuring reproducibility in stochastic algorithms.
    \item \code{n\_iter}: Number of iterations. The implemented algorithm lacks an intrinsic stopping rule and proceeds for a fixed number of iterations, independently of its convergence state. As a result, convergence must be evaluated a posteriori based on suitable diagnostic criteria (see section \ref{sec:ill_estimation}).
    \item \code{n\_burn\_in\_iter}: Number of iterations during the burn-in phase.
    \item \code{progress\_bar}: Used to display a progress bar of the iterations completed during computation.
\end{itemize}

\paragraph{What it outputs:}
Running \code{model.fit()} prints the name of the algorithm used and its total execution time. 
The model object which must be created beforehand (e.g., from a \pkg{leaspy} instance) is then updated during the fit. 
After fitting, the estimated population parameters are stored in \code{model.parameters} as a \proglang{Python} dictionary.
For the \code{LogisticModel}, the parameters are: \code{log_g_mean} ($\log g_k$), \code{log_v0_mean} ($\log v_k$), \code{betas_mean} ($\boldsymbol{\beta}$), \code{tau_mean} ($\overline{\tau}$), \code{tau_std} ($\sigma_\tau$), \code{xi_std} ($\sigma_\xi$) and \code{noise_std} ($\sigma_k$).

\subsection{Personalize} \label{sect:personalize}

\paragraph{What it does:}
In the personalization step, individual-specific parameters, the random effects, are estimated for each subject, conditioned on the previously estimated fixed effects. 
This step can be performed with different algorithms detailed in section~\ref{subsec: est_pred}.

\paragraph{Arguments:} The main arguments of interest are described below.
\begin{itemize}
    \item \code{data}: The dataset containing the subjects and their observations for personalization. It should follow \pkg{leaspy}'s formatting rules, that can either be the same data used during \code{fit} or a held-out test set.
    \item \code{algorithm}: Must be one of: \code{'scipy\_minimize'}, \code{'mean\_posterior'}, \code{'mode\_posterior'} as described in section~\ref{subsec: est_pred}.
     \item \code{seed}: If provided, the seed initializes the random number generator so it can produce the same sequence of numbers, ensuring reproducibility in stochastic personalization algorithms.
\end{itemize}

\paragraph{What it outputs:}
Running \code{model.personalize()} prints the name of the algorithm used and the total execution time. It returns an \code{IndividualParameters} object containing, for each subject, their individual random effects: \code{tau} ($\tau_i$), \code{xi} ($\xi_i$), and \code{sources} ($s_i$). The results can be accessed as a \code{pandas} DataFrame via \code{ip.to\_dataframe()}, or for a single subject via \code{ip["subject\_id"]}.

\subsection{Estimate}

\paragraph{What it does:}
The estimate function computes the modeled feature values for each subject, using the fixed effects and the subject's random effects.
This enables to reconstruct and predict individual progression. 
This is especially useful for plotting the progression of each feature, predicting new values, and filling missing observations.

\paragraph{Arguments:} The main arguments of interest are described below.
\begin{itemize}
    \item \code{timepoints}: a \code{pd.MultiIndex} or a \proglang{Python} dictionary mapping each subject ID to a list of time points (ages) at which to evaluate the trajectory.
    \item \code{individual\_parameters}: an \code{IndividualParameters} object containing the individual random effects obtained from \code{model.personalize()} or from an \code{IndividualParameters} object containing the individual random effects constructed from dictionary input to the \code{IndividualParameters} class.
    \item \code{to\_dataframe}: if \code{True}, it returns a \code{pd.DataFrame}; otherwise, it returns a dictionary mapping subject IDs to NumPy arrays.
\end{itemize}

\paragraph{What it outputs:}
\code{estimate} returns either a \code{pd.DataFrame} or a \proglang{Python} dictionary, containing the estimated value of each feature at each provided time point. 
For example, given 50 time points and 3 features, the result for each subject is an array of shape $(50 \times 3)$.

\subsection{Simulate} 
\label{sect:simulate}

\paragraph{What it does:} In \pkg{leaspy}, tools to generate synthetic longitudinal datasets from a fitted model are also provided. 
Such simulations are valuable for validation, benchmarking, and methodological research, particularly in medical domains where real data are highly sensitive and subject to privacy constraints that reduce their availability.

\paragraph{Arguments:} The main arguments of interest are described below. 
\begin{itemize}
    \item \code{algorithm}: The simulation algorithm to use. Currently, only one algorithm is implemented using the logistic model, so the value should be, by default, \code{simulate}.
    \item \code{features}: Outcomes of interest that the user wants to simulate. They must have been used to fit the model.
    \item \code{visit_parameters}: A \proglang{Python} dictionary specifying the visit schedule of the simulated patients, including:
    \begin{itemize}
        \item \code{patient_number}: Number of subjects to simulate
        \item \code{visit_type}: Three options are available:
        \begin{itemize}
            \item random spacing: Visit times and intervals are sampled from Gaussian distributions defined by \code{'random'}
            \item regular spacing: Visits occur at regular intervals defined by \code{'regular_visit'} 
            \item Custom spacing: Custom visit times are provided via a \code{pandas.DataFrame}
        \end{itemize}
    \end{itemize}
    \item \code{seed}: If provided, the seed initializes the random number generator so it can produce the same sequence of numbers, ensuring reproducibility in stochastic algorithms.
\end{itemize}

\paragraph{What it outputs:} The simulate function prints the execution time. 
It returns a \code{Result} class object containing: (i) the simulated individual parameters (\code{tau}, \code{xi}, \code{sources}) used to generate each virtual subject, (ii) the noise standard deviation, and (iii) a \code{Data} object from which a multi-indexed dataframe can be extracted, with the ID of each simulated subject and the time of each of their observations as the index, and one column per outcome.



\section{Illustration} \label{sec:illustrations}

In this section, we present an application of the modeling pipeline on a synthetic longitudinal dataset available directly in the \pkg{leaspy} package, so that all experiments presented here are fully reproducible. 
For a simple illustration, we use the logistic multivariate model. 
We walk through each step sequentially: loading and formatting longitudinal data, estimating population-level parameters via the \code{fit} procedure, deriving individual parameters through \code{personalize}, generating predictions with \code{estimate}, and finally simulating synthetic cohorts with \code{simulate}.
The results presented below are obtained in the reference Linux/AMD64 execution environment described in the computational details. 
On another architecture the stochastic MCMC-SAEM fit matches up to Monte Carlo variations.

\subsection{Context on the data}\label{sec:data_format}

We model the progression of Parkinson’s disease using synthetic data. 
The synthetic dataset contains repeated measurements for 200 subjects observed at multiple visits. 
We consider three clinical scores from the dataset: the MDS-UPDRS Part I \citep{goetz2008}, the SCOPA-AUT \citep{visser2004}, and the MoCA \citep{nasreddine2005}, all normalized to $[0, 1]$, with $0$ corresponding to no impairment and $1$ to maximum severity, as required by the logistic model. 
The following imports bring in the required modules and load the synthetic dataset from \pkg{leaspy}. 

\begin{CodeChunk}
\begin{CodeInput}
from leaspy.datasets import load_dataset
from leaspy.io.data import Data

df = load_dataset("parkinson")

\end{CodeInput}
\end{CodeChunk}

The dataset follows the long format expected by \pkg{leaspy}, indexed by subject identifier (\code{ID}) and visit age (\code{TIME}). 
Each row corresponds to one visit, and each column to one clinical score.

\begin{CodeChunk}
\begin{CodeInput}
print(df[FEATURES].head())
\end{CodeInput}
\begin{CodeOutput}
                  MDS1_total  SCOPA_total  MOCA_total
ID     TIME                                          
GS-001 71.354607    0.112301     0.160001    0.275257
       71.554604    0.140880     0.135852    0.380934
       72.054604    0.225499     0.211134    0.351172
       73.054604    0.132519     0.245323    0.377842
       73.554604    0.278923     0.223102    0.292768
\end{CodeOutput}
\end{CodeChunk}

We then check the number of subjects and split the dataset into a training set used to estimate population parameters, and a test set reserved for personalization and prediction.
\begin{CodeChunk}
\begin{CodeInput}
n_subjects = df.index.get_level_values("ID").unique().shape[0]
print(f"{n_subjects} subjects in the dataset.")
\end{CodeInput}
\begin{CodeOutput}
200 subjects in the dataset.
\end{CodeOutput}
\end{CodeChunk}

\begin{CodeChunk}
\begin{CodeInput}
df_train = df.loc[:"GS-160"][["MDS1_total", "SCOPA_total", "MOCA_total"]]
df_test  = df.loc["GS-161":][["MDS1_total", "SCOPA_total", "MOCA_total"]]
data_train = Data.from_dataframe(df_train)
data_test  = Data.from_dataframe(df_test)
\end{CodeInput}
\end{CodeChunk}

\begin{figure}[h]
    \centering
    \includegraphics[width=\linewidth]{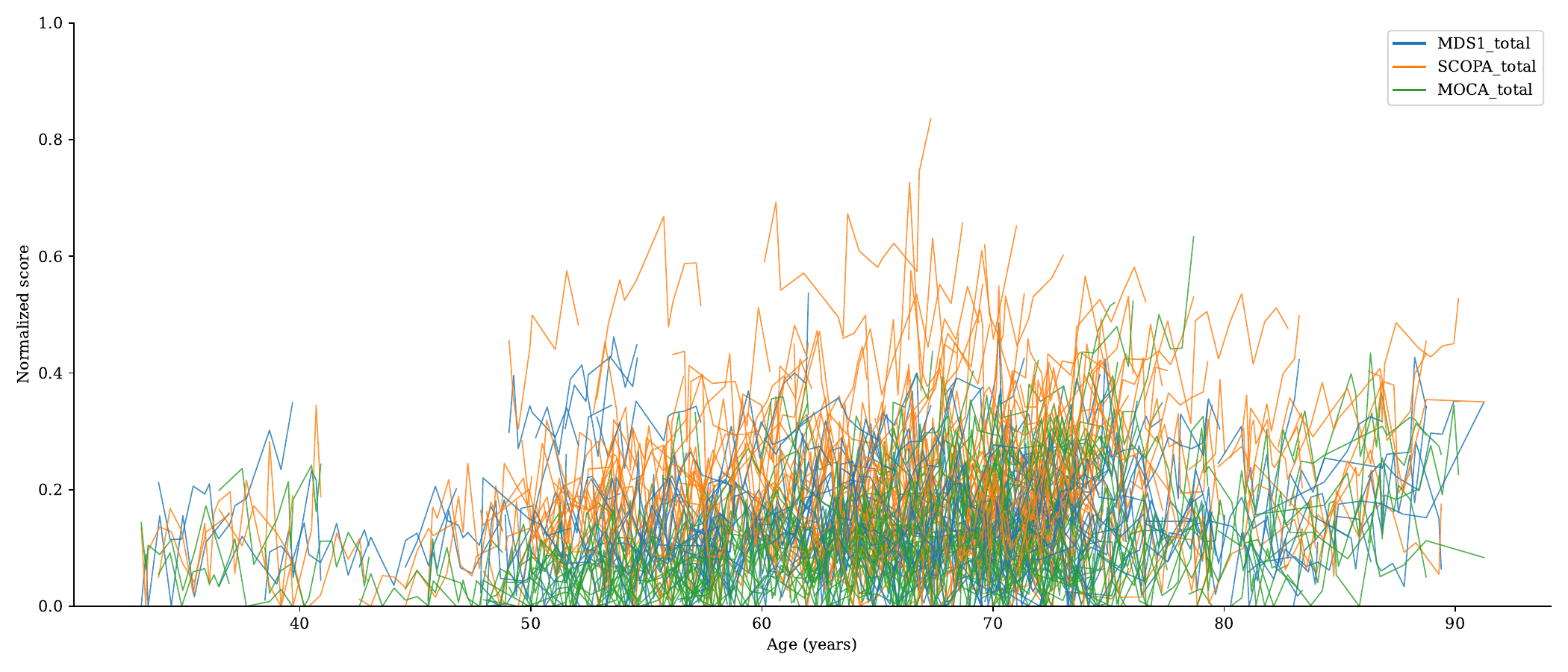}
    \caption{Raw observations for the three clinical scores across the 160 training subjects.}
    \label{fig:spaghetti}
\end{figure}

Figure~\ref{fig:spaghetti} displays individual trajectories of all training subjects in a spaghetti plot, plotted using standard \pkg{matplotlib} tools, and already reveals substantial inter-individual variability in both disease timing and progression speed.

\subsection{Model estimation}\label{sec:ill_estimation}

\paragraph{Model initialization:}
We initialize a multivariate logistic model with three features \\(\code{dimension=3}) and a two-dimensional source space, capturing two independent directions of inter-individual spatial variability reflecting the heterogeneity of Parkinson's disease progression across patients. 
The logistic model is further motivated by the bounded nature of the clinical scores, subject to floor and ceiling effects.

\begin{CodeChunk}
\begin{CodeInput}
from leaspy.models import LogisticModel
model = LogisticModel(name="logistic", dimension=3, source_dimension=2)
\end{CodeInput}
\end{CodeChunk}

\paragraph{Fit the model:}

The model is then fitted to the training data using the MCMC-SAEM algorithm. 
A fixed seed ensures reproducibility. 
The argument \code{n\_iter} controls the total number of iterations; here we run 100 000 iterations (approximately 33 minutes) to allow the stochastic approximations to stabilize. 
The arguments \code{save\_periodicity} and \code{plot\_periodicity} control how frequently the parameter estimates and their corresponding plots are written to disk during training (here every 500 iterations). 
The \code{path} argument specifies the output directory for these logs, and \code{overwrite\_logs\_folder} ensures the folder is reset if the run is repeated.

\begin{CodeChunk}
\begin{CodeInput}
model.fit(
    df_train, "mcmc_saem",
    seed=0, n_iter=100000, progress_bar=True,
    save_periodicity=500, plot_periodicity=500,
    path="_outputs/parkinson_fit_V21",
    overwrite_logs_folder=True,
)
\end{CodeInput}
\begin{CodeOutput}
Fit with `mcmc_saem` took: 32m 36.45s
\end{CodeOutput}
\end{CodeChunk}

\begin{figure}[h]
    \centering
    \includegraphics[width=\linewidth]{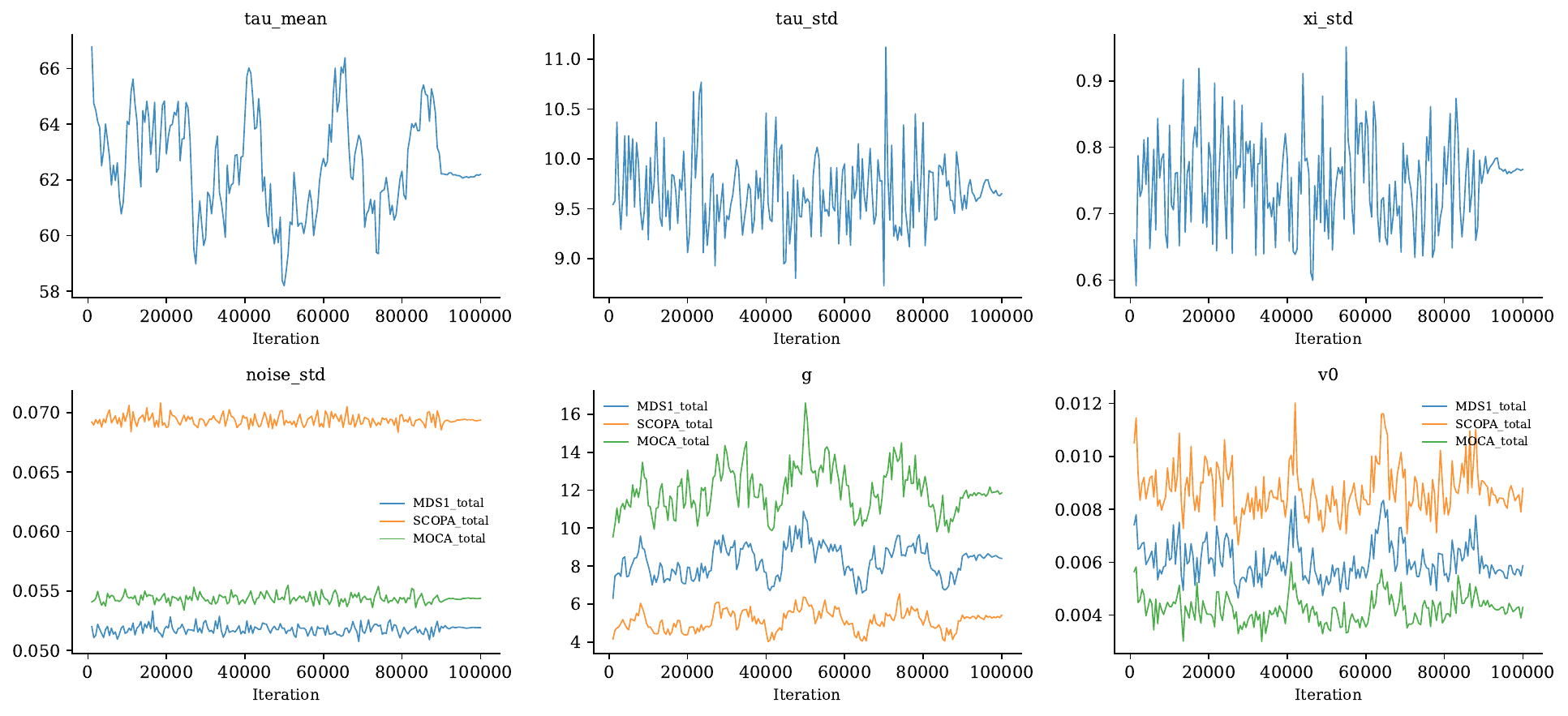}
    \caption{Evolution of key population parameters over the 100 000 MCMC-SAEM iterations (with 90,000 iterations for burn-in phase). Parameters \code{noise\_std}, \code{g} and \code{v0} stabilize rapidly, while \code{tau\_mean}, \code{tau\_std} and \code{xi\_std} exhibit larger but non-drifting fluctuations consistent with MCMC sampling noise.}
    \label{fig:convergence}
\end{figure}

\paragraph{Convergence assessment:}
Since the MCMC-SAEM algorithm does not implement an automatic stopping criterion, convergence is assessed visually by inspecting the traces of the estimated model parameters over iterations. 
The convergence traces of every model parameter are written to disk during the
fit, under the directory given by the \code{path} argument. Figure~\ref{fig:convergence}
is produced from those traces for the six parameters of interest: the population
reference time \code{tau_mean}, its standard deviation \code{tau_std}, the
individual log-acceleration standard deviation \code{xi_std}, the observation
noise \code{noise_std}, and the shape and velocity parameters \code{g} and
\code{v0}.
A satisfactory run is characterized by traces that stabilize around a plateau, with residual fluctuations attributable to the MCMC sampling noise.

\paragraph{Estimated parameters:}
Once convergence is established, a structured summary of the fitted model is accessible via \code{model.summary()}, which reports the estimated population parameters alongside model fit metrics and data context.

\begin{CodeChunk}
\begin{CodeInput}
model.summary()
\end{CodeInput}
\begin{CodeOutput}
================================================================================
                                 Model Summary                                  
================================================================================
Model Name: logistic
Model Type: LogisticModel
Features (3): MDS1_total, SCOPA_total, MOCA_total
Sources (2): Source 0 (s0), Source 1 (s1)
Observation Models: gaussian-diagonal
Neg. Log-Likelihood: -5592.5337
Parameters: 19
BIC: -13517.14
AIC: -13575.57

Training Metadata
--------------------------------------------------------------------------------
Algorithm: mcmc_saem
Seed: 0
Iterations: 100000

Data Context
--------------------------------------------------------------------------------
Subjects: 160
Visits: 1590
Total Observations: 4770
Leaspy Version: 2.1.0
================================================================================

Population Parameters
--------------------------------------------------------------------------------
  betas_mean:
                          s0        s1
            b0        0.0511    0.0038
            b1       -0.0232    0.0402
                    MDS1_to.  SCOPA_t.  MOCA_to.
  log_g_mean          2.1357    1.6673    2.4747
                    MDS1_to.  SCOPA_t.  MOCA_to.
  log_v0_mean        -5.1732   -4.7844   -5.4693

Individual Parameters
--------------------------------------------------------------------------------
  tau_mean           [62.1939]
  tau_std            [9.6497]
  xi_std             [0.7668]

Noise Model
--------------------------------------------------------------------------------
                    MDS1_to.  SCOPA_t.  MOCA_to.
  noise_std           0.0519    0.0694    0.0544

Derived Parameters (interpretable scale)
--------------------------------------------------------------------------------
                    MDS1_to.  SCOPA_t.  MOCA_to.
  v0                  0.0057    0.0084    0.0042
                    MDS1_to.  SCOPA_t.  MOCA_to.
  p0                  0.1057    0.1588    0.0776
================================================================================
\end{CodeOutput}
\end{CodeChunk}

\noindent The summary is organized in several blocks. The header reports model fit metrics: the negative log-likelihood, the number of estimated parameters, and the AIC and BIC criteria, which can be used for model selection and comparison. The training metadata block recalls the algorithm, seed, and number of iterations, ensuring reproducibility. The data context block summarizes the training set used.

The population parameters block reports the estimated fixed effects. The parameter \code{tau\_mean} gives the population reference age, here around 62 years. At this reference age, the derived parameters \code{p0} give the average score for each feature directly on the interpretable scale, here approximately $(0.1057, 0.1588, 0.0776)$ for the three clinical scores, and \code{v0} gives the corresponding progression speeds, approximately $(0.0057, 0.0084, 0.0042)$. These derived quantities are automatically computed from \code{log\_g\_mean} and \code{log\_v0\_mean} via $p_k = 1/(1 + \exp(\texttt{log\_g\_mean}_k))$ and $v_k = \exp(\texttt{log\_v0\_mean}_k)$, and are provided directly in the summary for interpretability.

The standard deviation \code{tau\_std} quantifies the spread of individual disease onsets around this reference (here approximately 10 years). The parameter \code{xi\_std} reflects the variability in progression speeds across the population: since individual log-accelerations $\xi_i$ are drawn from a distribution with standard deviation \code{xi\_std} = 0.767, a subject one standard deviation above the mean progresses at $\exp(0.767) \approx 2.2$ times the average speed. The \code{noise\_std} captures the residual observation noise for each feature after the model has accounted for inter-individual variability, here ranging from $0.0519$ to $0.0694$ across the three scores.

\begin{figure}[h]
    \centering
    \includegraphics[width=\linewidth]{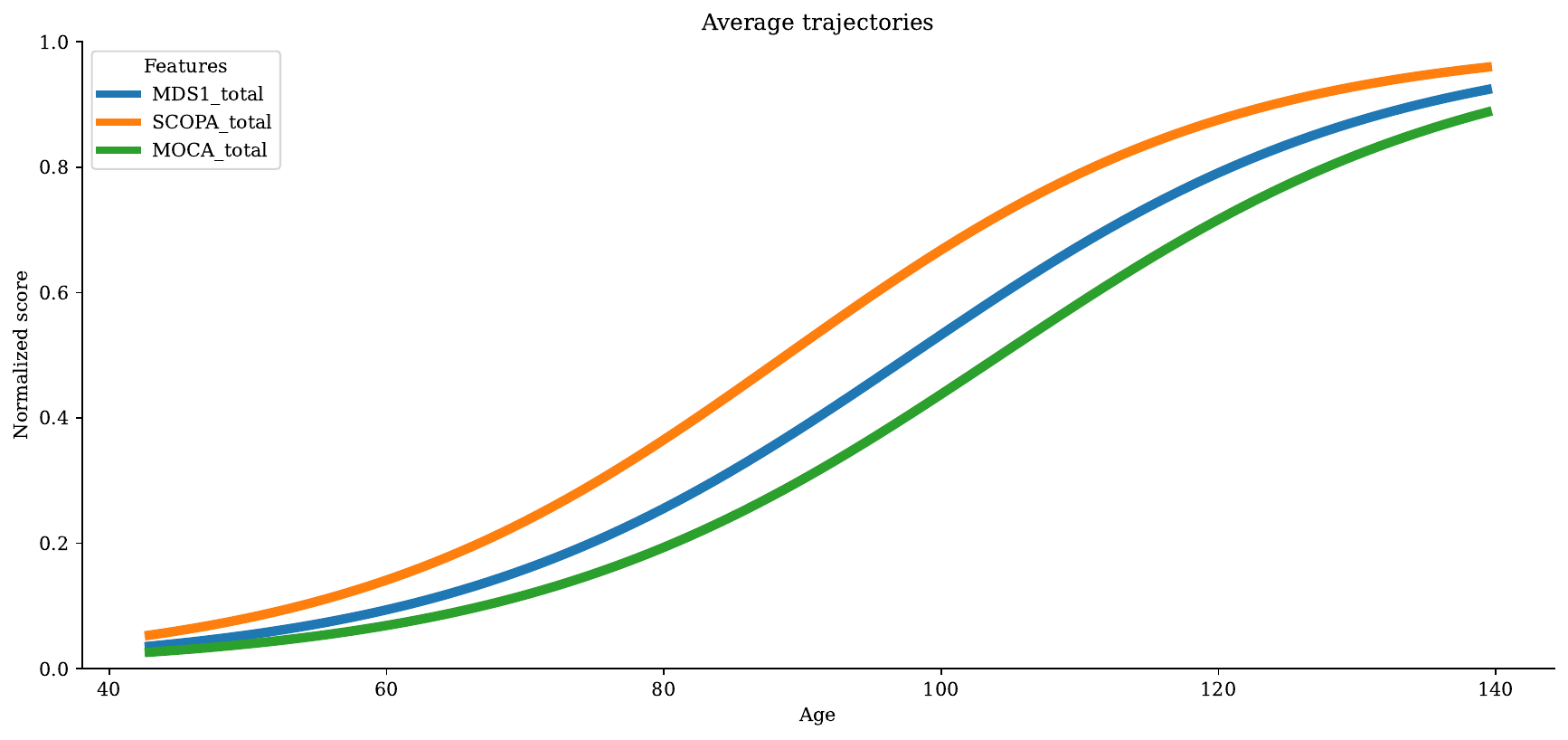}
    \caption{Population average trajectory plotted with \pkg{leaspy}. Each colored curve represents the expected progression of one clinical score as a function of reparametrized time.}
    \label{fig:average_trajectory_parkinson}
\end{figure}

\paragraph{Average trajectory:}
The estimated population trajectory, shown in Figure~\ref{fig:average_trajectory_parkinson}, corresponds to an individual with random effects set to their neutral values: $\tau_i = tau\_mean, \xi_i=0, \text{and } s_i=0$. 
It shows the average disease course, where each curve represents the expected normalized score as a function of the reparametrized time. 
This trajectory is obtained using \pkg{leaspy}'s built-in \code{Plotting} class.

\begin{CodeChunk}
\begin{CodeInput}
leaspy_plotting = Plotting(model)
ax = leaspy_plotting.average_trajectory(
    alpha=1, figsize=(14, 6), n_std_left=2, n_std_right=8
)
plt.savefig('figures/average_trajectory_V21.pdf', dpi=300, bbox_inches="tight")
plt.close("all")
\end{CodeInput}
\end{CodeChunk}

\subsection{Personalization and prediction}
\label{subsec:personalization}

The \code{personalize} function estimates individual-specific parameters for each subject, conditional on the fixed effects estimated during the fit. 
This yields, for each subject $i$, an estimate of the time-shift $\tau_i$, the log-acceleration $\xi_i$, and the sources $s_i$, which together fully characterize their individual trajectory. 
Here we use deterministic optimization via \code{scipy\_minimize}. 
Bayesian alternatives based on MCMC sampling are also available, as described in Section~\ref{sect:personalize}.

\paragraph{Reconstruction on the training set:}

We first assess the model's ability to reconstruct the training data by personalizing on \code{data_train} and comparing the reconstructed trajectories to the observed measurements.

\begin{CodeChunk}
\begin{CodeInput}
ip_train = model.personalize(data_train, "scipy_minimize", seed=0,
                            progress_bar=True)
\end{CodeInput}
\begin{CodeOutput}
Personalize with `scipy_minimize` took: 41.54s
\end{CodeOutput}
\end{CodeChunk}

\begin{figure}[h]
\centering
\includegraphics[width=\linewidth]{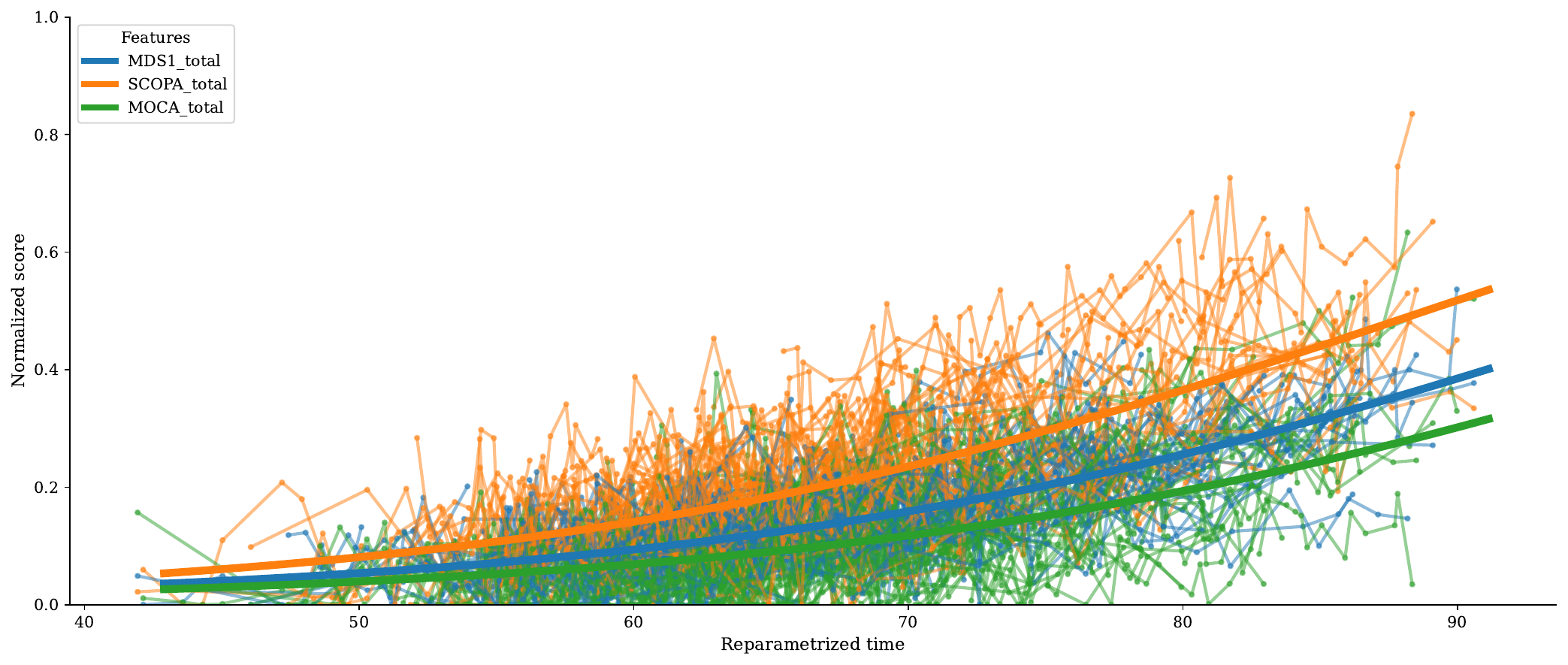}
\caption{Individual observations in reparametrized time overlaid with the estimated population trajectory for the three clinical scores.}
\label{fig:reparametrized}
\end{figure}

Thanks to the individual parameters estimated by \code{personalize}, observations can be plotted in reparametrized time using \pkg{leaspy}'s built-in visualization tools, which aligns all subjects to a common disease timescale. 
Figure~\ref{fig:reparametrized} overlays the individual observations in reparametrized time with the estimated population trajectory, showing that the model captures the overall progression trend across subjects.

\begin{CodeChunk}
\begin{CodeInput}
fig, ax = plt.subplots(figsize=(14, 6))
ax = leaspy_plotting.patient_observations_reparametrized(
    data_train, ip_train, alpha=0.5, linestyle="-", ax=ax
)
ax = leaspy_plotting.average_trajectory(
    alpha=1, n_std_left=2, n_std_right=3, ax=ax
)
ax.set_xlabel("Reparametrized time", fontsize=11)
ax.set_ylabel("Normalized score", fontsize=11)
ax.set_title("")
plt.tight_layout()
plt.savefig("figures/average_trajectory_with_reparametrized_obs.pdf",
            dpi=300, bbox_inches="tight")
plt.close("all")
\end{CodeInput}
\end{CodeChunk}

\noindent The resulting object \code{ip\_train} contains a dictionary with the estimated random effects for all subjects in \code{data\_train}. 
For instance, subject GS-001 has a time-shift $\tau = 68.58$, suggesting their clinical scores start deteriorating approximately 6.4 years later than the population average ($\tau_\text{mean} = 62.19$), and a log-acceleration $\xi = 1.45$ corresponding to a progression speed $\exp(1.45) \approx 4.3$ times the average.

\begin{CodeChunk}
\begin{CodeInput}
print(ip_train._individual_parameters["GS-001"])
\end{CodeInput}
\begin{CodeOutput}
{'sources': [-1.4963836669921875, 0.8069497346878052], 
     'tau': [68.58197784423828], 'xi': [1.4485920667648315]}
\end{CodeOutput}
\end{CodeChunk}

\begin{figure}[h]
\centering
\includegraphics[width=\linewidth]{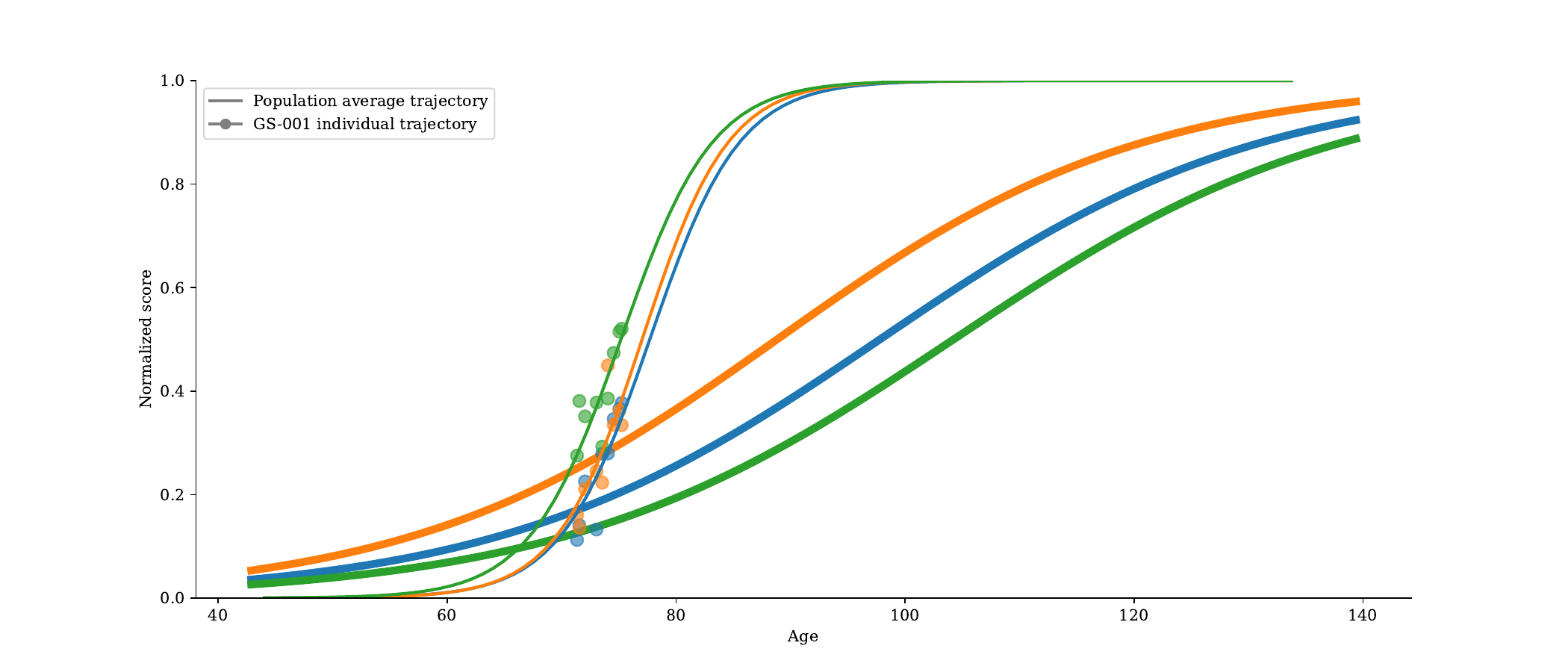}
\caption{Individual trajectory of subject GS-001 (thin curves with observations) overlaid with the population average trajectory (thick curves).}
\label{fig:individual_reconstruction}
\end{figure}

Individual trajectories can be visualized using the \code{patient_trajectories} method from \pkg{leaspy}'s visualization tools, overlaid here with the population average trajectory to illustrate the effect of the individual parameters. 
Figure~\ref{fig:individual_reconstruction} shows that subject GS-001 starts deteriorating later than the population average, consistent with their later disease onset, but progresses more steeply, consistent with their higher acceleration.

\begin{CodeChunk}
\begin{CodeInput}
fig, ax = plt.subplots(figsize=(14, 6))
ax = leaspy_plotting.average_trajectory(alpha=1, n_std_left=2, n_std_right=8, 
                                        ax=ax)
ax = leaspy_plotting.patient_trajectories(
    data_train, ip_train, patients_idx=["GS-001"],
    alpha=1, linestyle="-", linewidth=2, markersize=8, obs_alpha=0.6,
    figsize=(13, 5), factor_past=7, factor_future=15, ax=ax
)
ax.plot([], [], color="gray", linewidth=2,
        label="Population average trajectory")
ax.plot([], [], color="gray", linewidth=2, linestyle="-", marker="o",
        label="GS-001 individual trajectory")
ax.legend()
ax.set_title("")
plt.savefig('figures/individual_vs_average.pdf', dpi=800)
plt.close("all")
\end{CodeInput}
\end{CodeChunk}

Then, given the individual parameters estimated with \code{personalize}, the \code{estimate} function computes predicted feature values at any set of time points specified by the user, including time points beyond the observation window, a capability that will be exploited in the prediction section. 
Here we first use it to reconstruct trajectories at the observed visit times of the training set, which allows direct comparison with the measurements.

To quantify reconstruction quality, we compute the MAE and $R^2$ for each feature using \pkg{scikit-learn}.

\begin{CodeChunk}
\begin{CodeInput}
from sklearn.metrics import mean_absolute_error, r2_score
import statsmodels.api as sm
reconstructed = model.estimate(df_train.index, ip_train)
for feature in FEATURES:
    mae = mean_absolute_error(df_train[feature], reconstructed[feature])
    r2  = r2_score(df_train[feature], reconstructed[feature])
    print(f"{feature}: MAE={mae:.3f}, R2={r2:.3f}")
\end{CodeInput}
\begin{CodeOutput}
MDS1_total: MAE=0.039, R2=0.717
SCOPA_total: MAE=0.052, R2=0.760
MOCA_total: MAE=0.042, R2=0.725
\end{CodeOutput}
\end{CodeChunk}

\begin{figure}[h]
\centering
\includegraphics[width=\linewidth]{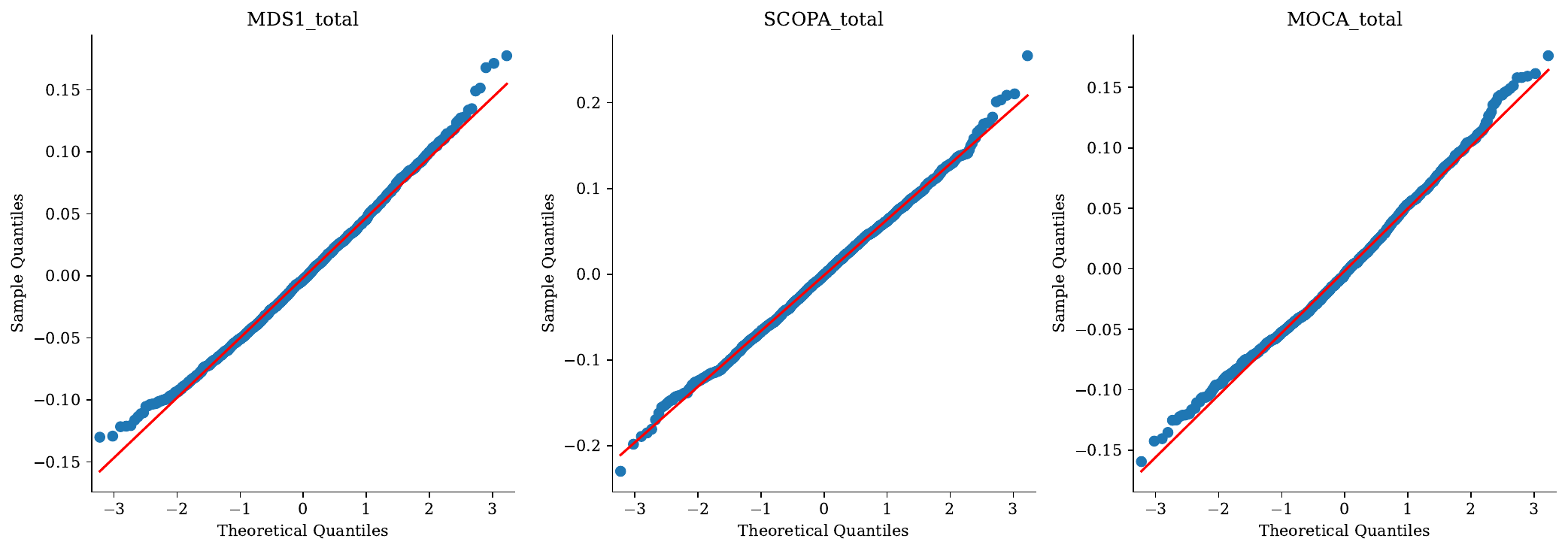}
\caption{Residual Q-Q plots for the three clinical scores on the training set. 
Points align closely with the reference line across all features, supporting the normality assumption on the observation noise.}
\label{fig:qqplot}
\end{figure}

Reconstruction errors are low across all three features (MAE between 0.039 and 0.052), consistent with the estimated observation noise \code{noise\_std}. The model explains approximately 72-76\% of the observed variance ($R^2$ between 0.717 and 0.760). 
The residual Q-Q plots (Figure~\ref{fig:qqplot}) confirm that the normality assumption on the observation noise is well satisfied, with only minor deviations at the tails.

\begin{CodeChunk}
\begin{CodeInput}
fig, axes = plt.subplots(1, 3, figsize=(14, 5))
for ax, feature in zip(axes, FEATURES):
    residuals = df_train[feature].values - reconstructed[feature].values
    sm.qqplot(residuals, line="s", ax=ax)
    ax.set_title(feature)
plt.tight_layout()
plt.savefig("figures/qqplot_reconstruction.pdf", dpi=300,
            bbox_inches="tight")
plt.close(fig)
\end{CodeInput}
\end{CodeChunk}

\paragraph{Prediction on the test set:}

We now evaluate the model's ability to predict future observations for new subjects. 
For each subject in the test set, we personalize the model on all visits except the last one, and then predict the score at the last visit time, simulating a realistic clinical scenario where one aims to anticipate future disease progression from partial follow-up data.

\begin{CodeChunk}
\begin{CodeInput}
df_test_partial = (
    df_test.groupby("ID").apply(lambda x: x.iloc[:-1]).droplevel(0)
)
df_test_last = df_test.groupby("ID").tail(1)
data_test_partial = Data.from_dataframe(df_test_partial)
ip_test = model.personalize(data_test_partial, "scipy_minimize", seed=0, 
                            progress_bar=True)
\end{CodeInput}
\begin{CodeOutput}
Personalize with `scipy_minimize` took: 10.24s
\end{CodeOutput}
\end{CodeChunk}

The individual trajectories estimated from partial follow-up can be visualized using the \code{patient\_trajectories} method from \pkg{leaspy}'s visualization tools, which extrapolates the fitted trajectory beyond the last observed visit.

\begin{CodeChunk}
\begin{CodeInput}
import matplotlib.image as mpimg
fig, axes = plt.subplots(1, 2, figsize=(18, 5))
fig.suptitle("Observations and individual trajectories", fontsize=13)
for ax, patient_id in zip(axes, ["GS-170", "GS-175"]):
    tmp_path = f"figures/tmp_{patient_id}.png"
    ax_ind = leaspy_plotting.patient_trajectories(
        data_test_partial, ip_test, patients_idx=[patient_id],
        alpha=1, linestyle="-", linewidth=2, markersize=8,
        obs_alpha=0.6, figsize=(9, 5), factor_past=0.5, factor_future=5,
    )
    ax_ind.set_title("")
    ax_ind.set_xlim(63, 90)
    ax_ind.get_figure().savefig(tmp_path, dpi=300, bbox_inches="tight")
    plt.close(ax_ind.get_figure())
    ax.imshow(mpimg.imread(tmp_path))
    ax.axis("off")
    ax.set_title(f"Patient: {patient_id}", fontsize=11)
plt.tight_layout()
plt.savefig('figures/individual_trajectories.pdf', dpi=300,
            bbox_inches="tight")
plt.close("all")
\end{CodeInput}
\end{CodeChunk}

\begin{figure}[h]
    \centering
    \includegraphics[width=\linewidth]{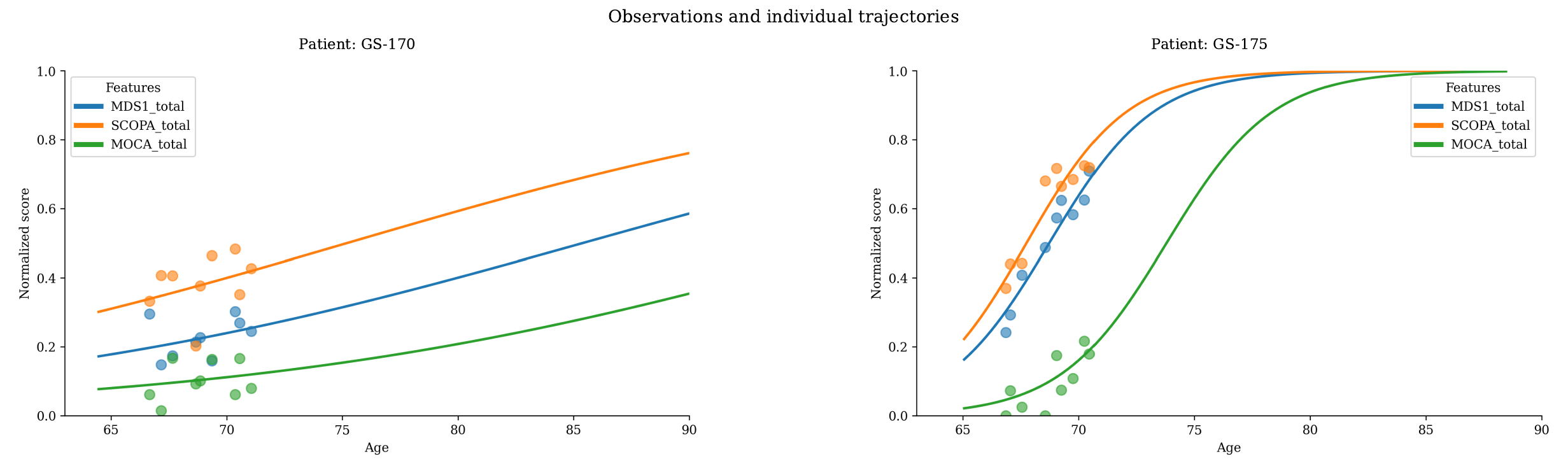}
    \caption{Estimated individual trajectories for two representative test subjects. Dots represent observed measurements; curves show the fitted logistic trajectory extrapolated beyond the last visit.}
    \label{fig:individual_trajectories}
\end{figure}

Figure~\ref{fig:individual_trajectories} shows two representative subjects from the test set. 
Subject GS-170 is at an early disease stage: all three scores remain relatively low throughout the observation window, and the model extrapolates a gradual progression beyond the last visit.
Subject GS-175 illustrates a more advanced profile: \code{MDS1\_total} and \code{SCOPA\_total} have already entered the steep part of their logistic curve at the time of observation, while \code{MOCA\_total} lags substantially behind, reflecting a different ordering of symptom progression captured by the spatial random effects.

The \code{estimate} function can compute predicted value of each feature at any set of time points, including time points beyond the observation window. 
For instance, for subject GS-170, whose visits were recorded between ages 66.7 and 71.6, one can request predicted scores between ages 72 and 80:

\begin{CodeChunk}
\begin{CodeInput}
observations = df.loc["GS-170"]
print(f"Observed ages: {observations.index.values.round(1)}")
\end{CodeInput}
\begin{CodeOutput}
Observed ages: [66.7 67.2 67.7 68.7 68.9 69.4 70.4 70.6 71.1 71.6]
\end{CodeOutput}
\end{CodeChunk}

\begin{CodeChunk}
\begin{CodeInput}
timepoints = np.linspace(72, 80, 10)
predicted = model.estimate({"GS-170": timepoints}, ip_test)
\end{CodeInput}
\end{CodeChunk}

\noindent The resulting object \code{predicted} is a dictionary mapping each subject identifier to an array of predicted feature values at the requested time points. 

Beyond visualization, \code{estimate} can serve as the basis for quantitative prediction evaluation by comparing predicted and observed values at held-out time points. 
Here we evaluate the model at the last visit time of each test subject (withheld during personalization) using \pkg{scikit-learn} to compute MAE and $R^2$. 

\begin{CodeChunk}
\begin{CodeInput}
predicted_last = model.estimate(df_test_last.index, ip_test)
for feature in FEATURES:
    mae = mean_absolute_error(df_test_last[feature], predicted_last[feature])
    r2  = r2_score(df_test_last[feature], predicted_last[feature])
    print(f"{feature}: MAE={mae:.3f}, R2={r2:.3f}")
\end{CodeInput}
\begin{CodeOutput}
MDS1_total: MAE=0.047, R2=0.761
SCOPA_total: MAE=0.060, R2=0.741
MOCA_total: MAE=0.047, R2=0.535
\end{CodeOutput}
\end{CodeChunk}

Prediction errors remain low across all three features (MAE between $0.047$ and $0.060$), with $R^2$ values ranging from $0.535$ to $0.761$, indicating that the model generalizes well to unseen subjects despite being fitted on a separate training cohort. 
The lower $R^2$ for \code{MOCA\_total} suggests greater inter-individual variability in cognitive progression that the model only partially captures.

The evaluation procedure presented above, where individual parameters are estimated from all but the last visit and the held-out visit is predicted and compared to the ground truth for every subject in the test set, constitutes a form of cross-validation adapted to longitudinal data, providing an unbiased estimate of the model's predictive performance without requiring the population parameters to be re-estimated for each subject.

\subsection{Data simulation}

The \code{simulate} function generates a synthetic longitudinal dataset from a previously fitted model, by sampling individual parameters $(\tau_i, \xi_i, s_i)$ from the population distributions estimated during the fit and simulating observations at visit times defined by a user-specified schedule.

The visit schedule is controlled through a \proglang{Python} dictionary (i.e., a mapping from parameter names to their corresponding values) of parameters: \code{patient\_number} sets the number of synthetic subjects to generate, while \code{visit\_type} specifies how visit times are sampled. Here we use \code{visit\_type = "random"}, where visit times and intervals are sampled from Gaussian distributions, but other scheduling options are available, as described in Section~\ref{sect:simulate}.

\begin{CodeChunk}
\begin{CodeInput}
visits_params = {
    "patient_number": 200,
    "visit_type": "random",
    "first_visit_mean": 0.0,
    "first_visit_std": 0.4,
    "time_follow_up_mean": 11,
    "time_follow_up_std": 0.5,
    "distance_visit_mean": 2 / 12,
    "distance_visit_std": 0.75 / 12,
    "min_spacing_between_visits": 1 / 365,
}

simulated = model.simulate(
    algorithm="simulate",
    features=FEATURES,
    visit_parameters=visits_params,
    seed=SEED,
)
df_sim = simulated.data.to_dataframe().set_index(["ID", "TIME"])
print(df_sim.head())
\end{CodeInput}
\noindent The output follows the same long format as the original dataset, indexed by subject identifier and visit age, and can therefore be used directly as input to \code{fit} or any downstream analysis.
\begin{CodeOutput}
Simulate with `simulate` took: 0.67s
           MDS1_total  SCOPA_total  MOCA_total
ID TIME
0  59.196    0.178540     0.070239    0.028089
   59.306    0.150122     0.210433    0.169825
   59.350    0.182084     0.357574    0.051757
   59.616    0.267059     0.177252    0.045282
   59.823    0.215180     0.134273    0.043163

\end{CodeOutput}
\end{CodeChunk}



To assess whether the simulated data preserve the structure of the original fitted model, we fitted a new \code{LogisticModel} on the synthetic cohort and compared its parameters with those of the model used to generate the data. In the following outputs, empty entries in the \code{Feature} column correspond to parameters that are not feature-specific.

\begin{CodeChunk}
    \begin{CodeInput}
sim_model = LogisticModel(name="logistic", dimension=3, source_dimension=2)
sim_model.fit(
    df_sim, "mcmc_saem",
    seed=SEED, n_iter=100000, progress_bar=True,
    save_periodicity=500, plot_periodicity=500,
    path="_outputs/parkinson_simulation_fit_V21",
    overwrite_logs_folder=True,
)
    \end{CodeInput}
    \begin{CodeOutput}
Fit with `mcmc_saem` took: 59m 49.79s
\end{CodeOutput}
\end{CodeChunk}

We compute how the refitted model recovers the population trajectory parameters.

\begin{CodeChunk}
\begin{CodeInput}
rows = []
for param in ["tau_mean", "tau_std", "xi_std", 
                "noise_std", "log_g_mean", "log_v0_mean"]:
    true = model.parameters[param].detach().cpu().numpy().ravel()
    est = sim_model.parameters[param].detach().cpu().numpy().ravel()
    if param in ["log_g_mean", "log_v0_mean"]:
        true, est = np.exp(true), np.exp(est)

    parameter = {"log_g_mean": "g", "log_v0_mean": "v0"}.get(param, param)
    features = FEATURES if len(true) == len(FEATURES) else [""]

    for feature, true_value, est_value in zip(features, true, est):
        rows.append({
            "Parameter": parameter,
            "Feature": feature,
            "Original": round(float(true_value), 5),
            "Refit": round(float(est_value), 5),
            "Relative error (%)": round(
                float((est_value - true_value) / abs(true_value) * 100), 2
                ),
        })

population_recovery = pd.DataFrame(rows)
print(population_recovery.to_string(index=False))
\end{CodeInput}
\begin{CodeOutput}
Parameter     Feature  Original    Refit  Relative error (%)
 tau_mean              62.19386 61.16427               -1.66
  tau_std               9.64974  8.79395               -8.87
   xi_std               0.76675  0.89134               16.25
noise_std  MDS1_total   0.05192  0.05134               -1.13
noise_std SCOPA_total   0.06938  0.06992                0.78
noise_std  MOCA_total   0.05439  0.05338               -1.84
        g  MDS1_total   8.46336  8.07922               -4.54
        g SCOPA_total   5.29758  5.13110               -3.14
        g  MOCA_total  11.87867 11.55059               -2.76
       v0  MDS1_total   0.00567  0.00570                0.52
       v0 SCOPA_total   0.00836  0.00844                0.98
       v0  MOCA_total   0.00421  0.00416               -1.25
\end{CodeOutput}
\end{CodeChunk}

The reference time $\overline{\tau}$ differs by less than $2\%$, same for the observation noise parameters $\sigma_k$, and the feature-specific parameters $g_k$ and $v_k$ remain within approximately $5\%$ of the generating values. The largest discrepancies concern the random-effect standard deviations, in particular $\sigma_\xi$ ($16.25\%$), which is expected since these quantities depend directly on the sampled distribution of individual effects in a finite simulated cohort. This finite-sample variability at the population level does not prevent an accurate recovery of the individual random effects, as assessed below.

At the individual level, the parameters sampled during simulation provide a reference against which the parameters estimated by personalization can be compared. We personalized the refitted model on the simulated data and compared the true and estimated values of the temporal random effects $\xi_i$ and $\tau_i$, as well as the feature-specific spatial shifts $w_{ik}$.

\begin{CodeChunk}
\begin{CodeInput}
data_sim = Data.from_dataframe(df_sim)
ip_sim_est = sim_model.personalize(
    data_sim, "scipy_minimize", seed=SEED, progress_bar=True
)

ip_true = simulated.individual_parameters.astype(float)
ip_est = ip_sim_est.to_dataframe().astype(float)

w_cols = [f"w_{feature}" for feature in FEATURES]
source_cols = [f"sources_{i}" for i in range(sim_model.source_dimension)]
mixing_matrix = sim_model.state.get_tensor_value(
    "mixing_matrix"
).detach().cpu().numpy()

true_w = ip_true[
    [f"w_{i}" for i in range(len(FEATURES))]
].set_axis(w_cols, axis=1)
est_w = pd.DataFrame(
    ip_est[source_cols].to_numpy() @ mixing_matrix, 
    index=ip_est.index, columns=w_cols
)
ip_true = pd.concat([ip_true[["xi", "tau"]], true_w], axis=1)
ip_est = pd.concat([ip_est[["xi", "tau"]], est_w], axis=1)

columns = [("xi", r"$\xi_i$", "", "xi"), ("tau", r"$\tau_i$", "", "tau")] + [
    (f"w_{feature}", rf"$\text{{w}}_{{i{k}}}$ ({feature})", feature, "w")
    for k, feature in enumerate(FEATURES)
]


def icc_3_1(x, y):
    values = np.c_[x, y]
    row_means = values.mean(axis=1)
    ms_between = 2 * ((row_means - values.mean()) ** 2).sum() / (len(x) - 1)
    ms_error = ((values - row_means[:, None]) ** 2).sum() / len(x)
    return (ms_between - ms_error) / (ms_between + ms_error)


common_ids = ip_true.index.intersection(ip_est.index)
score_rows = []
for column, _, feature, parameter in columns:
    x = ip_true.loc[common_ids, column].to_numpy()
    y = ip_est.loc[common_ids, column].to_numpy()
    score_rows.append({
        "Parameter": parameter,
        "Feature": feature,
        "ICC(3,1)": round(float(icc_3_1(x, y)), 3),
        "Pearson r": round(float(np.corrcoef(x, y)[0, 1]), 3),
    })

individual_recovery = pd.DataFrame(score_rows)
print(individual_recovery.to_string(index=False))
\end{CodeInput}
\begin{CodeOutput}
Personalize with `scipy_minimize` took: 53.04s
Parameter     Feature  ICC(3,1)  Pearson r
       xi                 0.965      0.967
      tau                 0.979      0.983
        w  MDS1_total     0.989      0.991
        w SCOPA_total     0.992      0.992
        w  MOCA_total     0.992      0.993
\end{CodeOutput}
\end{CodeChunk} 

\begin{figure}[h]
\centering
\includegraphics[width=0.85\linewidth]{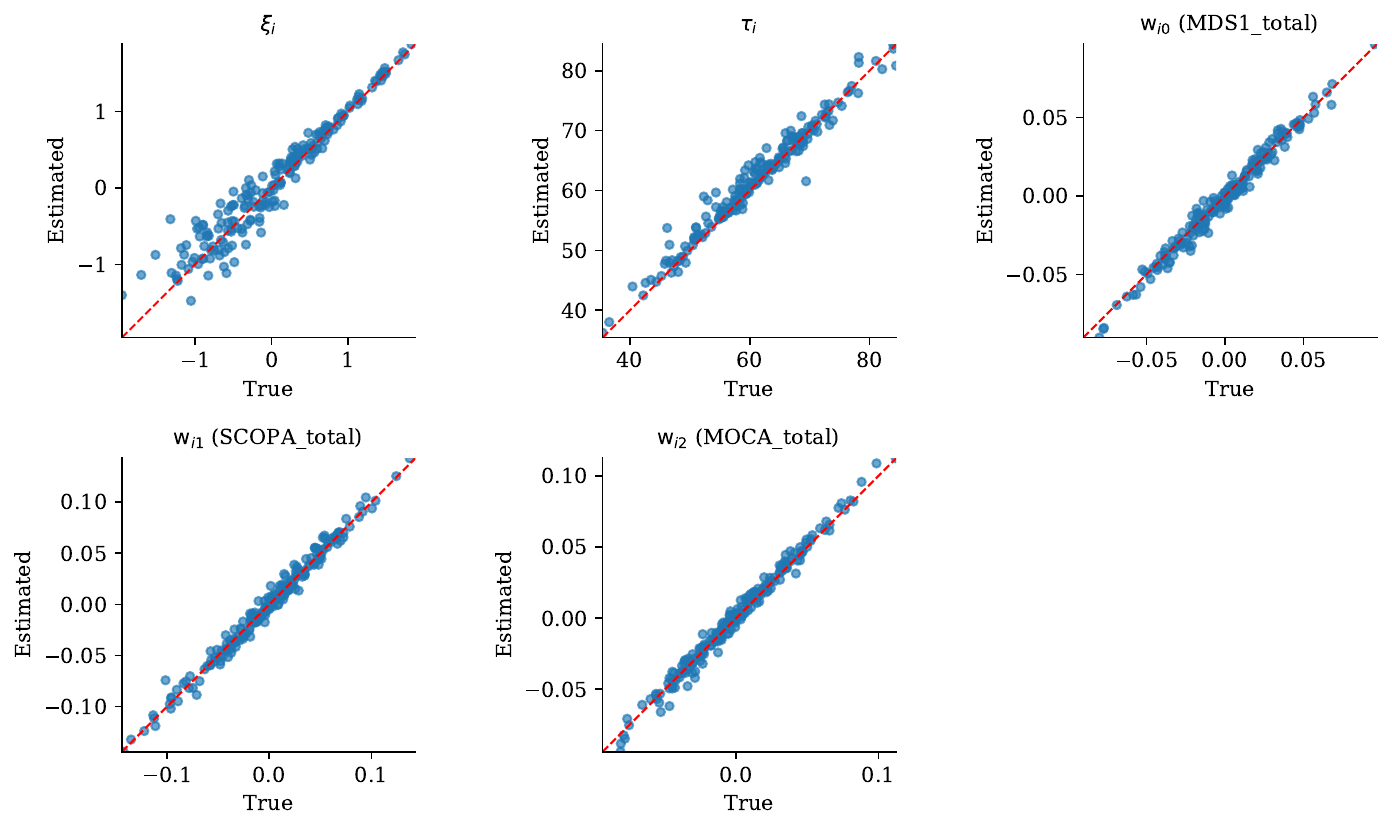}
\caption{Recovery of individual-level parameters in the simulated cohort. Each point corresponds to one simulated subject; the dashed red line indicates perfect agreement between the true simulated parameter and the personalized estimate.}
\label{fig:simulation_individual_recovery}
\end{figure}

The agreement is high for all individual parameters, with ICC values above $0.95$. The spatial shifts $w_{ik}$ are recovered particularly well, with correlations close to $0.99$ for all three clinical scores. Figure~\ref{fig:simulation_individual_recovery} confirms this result visually: the estimated parameters lie close to the identity line, with no apparent systematic bias. Together, these results indicate that \code{simulate} generates synthetic observations from which both the population-level structure and the subject-specific variability of the original fitted model can be recovered.

\begin{CodeChunk}
\begin{CodeInput}
fig, axes = plt.subplots(3, 3, figsize=(10, 8))
axes = axes.ravel()  

for ax, (column, label, _, _) in zip(axes, columns):
    x = ip_true.loc[common_ids, column]
    y = ip_est.loc[common_ids, column]
    low = min(x.min(), y.min())
    high = max(x.max(), y.max())
    ax.scatter(x, y, s=16, alpha=0.65)
    ax.plot([low, high], [low, high], "r--", lw=1)
    ax.set_xlim(low, high)
    ax.set_ylim(low, high)
    ax.set_aspect("equal", adjustable="box")
    ax.set_title(label, fontsize=10)
    ax.set_xlabel("True")
    ax.set_ylabel("Estimated")

for ax in axes[len(columns):]:
    ax.set_visible(False)

plt.tight_layout()
plt.savefig(
    "figures/simulation_individual_true_vs_est.pdf", dpi=300, 
    bbox_inches="tight"
)
plt.close("all")       
\end{CodeInput}
\end{CodeChunk}

\section{Discussion} \label{sec:discussion}

In this paper, we have introduced the \pkg{leaspy} library, a user-friendly \proglang{Python} progression modeling tool.
It has been designed with extensibility in mind, enabling the integration of new models and algorithms as scientific and methodological advances emerge. 
It implements a mixed-effects model with time reparametrization that aligns individuals along a common latent timeline. 
One of the major challenges in the study of complex evolving systems, such as economic market lifecycles, mechanical component degradation and chronic disease progression is to accurately quantify trajectory dynamics while accounting for the substantial individual heterogeneity. 
The model can jointly handle multivariate variables while explicitly capturing their inter-feature relationships, rather than considering them as a simple aggregation of independent measures. 
Modern empirical research aimed at uncovering the structural mechanisms underlying these conditions often relies on repeated measurements collected over extended periods of time. 
As a result, progression models must be able to analyze complex longitudinal data in a principled manner.
By positioning \pkg{leaspy} within the broader landscape of nonlinear mixed-effects modeling software, this work highlights the framework's ability to bridge the gap between rigorous MCMC estimation and accessible, object-oriented software engineering.
Unlike traditional tools that require manual hardcoding or rely on analytical approximations, \pkg{leaspy} integrates advanced statistical methods with modern software standards ensuring usability even for non-expert users.

A key advantage of the proposed model is that it explicitly incorporates time reparametrization, mapping observation tile to a latent age, allowing direct comparisons on a shared timeline. 
This allows meaningful comparisons to be made between subjects at similar points in their trajectory, regardless of differences their actual observation times or onset timepoints. 
Additionally, the model has a multivariate formulation, in which each score is associated with score-specific parameters that define its trajectory. 
This structure allows multiple outcomes to be modeled jointly with a common parametrization, rather than as independent univariate trajectories, ensuring that the overall progression is smooth and coherent across scores. 
Score-specific shifts indicate which outcomes are more advanced relative to the population trajectory, further enhancing interpretability.
Also, parameter estimation is performed using the MCMC-SAEM algorithm, which efficiently handles nonlinear mixed-effects models with complex latent random effects, providing stable and reliable estimates compared with deterministic optimization methods~(\cite{Panhard2008}).
Moreover, by using population priors as a regularization tool, \pkg{leaspy} stabilizes the estimation process and prevents the model from overfitting or becoming unidentifiable. 
This design choice ensures numerical stability during the SAEM algorithm while clearly separating group-level trends from individual noise.

The \pkg{leaspy} library has been developed to closely follow the formulation of the Disease Course Mapping model. 
It is an open-source library, available on \href{https://github.com/aramis-lab/leaspy}{\textcolor{blue}{\code{GitHub}}}, designed to promote reproducibility and facilitate continuous integration and continuous development (CI/CD). 
Automated testing and CI pipelines ensure that updates to the library do not break existing functionality, while version control and open access enable other researchers to reproduce analyses and extend the software with new models or datasets. 
By adhering to these practices, \pkg{leaspy} provides a stable and reliable framework for disease progression modeling, supporting both methodological transparency and collaborative development.
In the latest released version 2.1.0, several models are implemented, allowing the modeling of continuous data using logistic or linear trajectories, the joint modeling of longitudinal and censored data, and a mixture model that allows the unsupervised clustering of individuals according to their progression profiles.
At the same time, as demonstrated in Section \ref{sec:illustrations}, its straightforward interface allows users with moderate \proglang{Python} experience to easily analyze and interpret complex longitudinal data. 
The library is designed to be accessible to a wide range of researchers, helping bridge the gap between statistical, programming, and clinical expertise.
Moreover, it comes with a comprehensive and well-organized \href{https://leaspy.readthedocs.io/en/latest/}{\textcolor{blue}{documentation page}}, providing detailed explanations, usage examples, and guidance for all its functions.

The library comes with some limitations. 
Currently, the main diagnostic tool available for model evaluation is the convergence plots, which may restrict the ability to fully assess model performance in more complex or nuanced scenarios. 
While model selection criteria such as AIC and BIC have been incorporated into the latest release, additional diagnostic tools for model evaluation, such as residual analysis and posterior predictive checks, are not yet included in the library.
In the meantime, users may rely on other libraries until native support is added in future releases.
As part of our commitment to continuous improvement, several models are under development. 
For example, discrete data, covariance structures, and non-monotonic trajectories are areas where preliminary work has been done, but fully implemented models are not yet available in the latest released version. 
Also, data preprocessing, including normalization and standardization, must currently be handled by the user, which can introduce variability and extra effort.

We envision \pkg{leaspy} as a collaborative and evolving tool that not only supports reproducible research but also accelerates progress in progression modeling and longitudinal data analysis. 
By continuously incorporating user feedback, expanding model coverage, and providing comprehensive documentation and practical examples, the library aims to empower researchers from diverse backgrounds to explore complex longitudinal data with confidence and efficiency, bridging the gap between statisticians, programming engineers, and applied researchers.


\section*{Computational details}


The reference results in this paper were obtained on Linux/AMD64 with an AVX2-capable CPU, using
\proglang{Python}~3.10.19, CPU-only \pkg{torch}~2.7.1, and
\pkg{leaspy}~2.1.0, which is distributed under a BSD-3-Clause-Clear license.
Deterministic \pkg{torch} algorithms and a single intra-op and inter-op thread were used, with \code{PYTHONHASHSEED=0}.
The code accompanying this article includes a \code{Dockerfile} that builds this
environment, fixing the operating system, the \proglang{Python} version, every
dependency version, and the CPU instruction set used by \pkg{torch}. Within it
the workflow is bitwise reproducible: eight complete runs on five machines,
spanning three Intel generations and two Apple Silicon generations, produced
identical model parameters, tabulated results and figures. Outside it, results
may differ in the final digits, as numerical libraries and CPU instruction sets
differ between platforms, while the estimated parameters and the conclusions
drawn from them are unchanged.

\section*{Acknowledgments}

\begin{leftbar}
Juliette Ortholand and Sofia Kaisaridi are co-first authors and contributed equally to this work.
Sophie Tezenas du Montcel is the corresponding author.
The research leading to these results has received funding: from the program "Investissements d’avenir" by the French government under management of Agence Nationale de la Recherche (reference ANR-10-IAIHU-06, ANR-19-P3IA-0001 (PRAIRIE 3IA Institute), ANR-19-JPW2-000 (E-DADS)); by H2020 programme (project EuroPOND (666992), project HBP SGA1 (720270), project TVB-Cloud (826421)); by the European Research Council (project LEASP (678304)); by the ICM Big Brain Theory Program (project DYNAMO); by the REWIND project (pRecision mEdicine WIth loNgitudinal Data), as part of the PEPR Santé Numérique (grant number 2023000354-03).
\end{leftbar}


\bibliography{refs}


\newpage

\end{document}